\documentclass[journal]{vgtc}                     

\onlineid{0}

\vgtccategory{Research}

\title{Syndesmoscope: The Power of Invariant Plots\\Linked to Traditional Network Views}

\author{%
  \authororcid{Matt I. B. Oddo}{0009-0009-3187-4994},
  \authororcid{Indira Sowy}{0009-0005-4563-7165},
  \authororcid{Stephen Kobourov}{0000-0002-0477-2724}, and
  \authororcid{Tamara Munzner}{0000-0002-3294-3869}
}

\authorfooter{
  \item
        Matt Oddo, Indira Sowy, and Tamara Munzner are with the Department of Computer Science, University of British Columbia. Email: \{bofarull,iasowy,tmm\}@cs.ubc.ca
  \item
        Stephen Kobourov is with the Department of Computer Science at the Technical University of Munich. Email: stephen.kobourov@tum.de
}

\abstract{Traditional network representations, such as node-link views and adjacency matrices, can show dramatically different visual patterns, depending on the underlying layout or seriation algorithm. In contrast, invariant plots consistently surface the same visual pattern for the same input topology; yet researchers have underexplored them and have not integrated them into visualization systems. We present Syndesmoscope, an interactive system for network exploration that juxtaposes multiple views of the same network. Panes show a familiar a force-directed view alongside three panes with interpretable geometric layouts based on graph-theoretic properties: dense-sparse gradient, geodesic eccentricity, and spectral bisection. As a secondary contribution, we introduce kSnakes, a new invariant plot based on density decomposition. Syndesmoscope supports two key interactions: leapfrogging, or linked highlighting between different and interpretable visual patterns; and hopscotching, or hop-based traversal that extends data selections through the underlying topology. Through usage scenarios across a corpus of 72 diverse networks, we demonstrate how these interactions reveal network patterns inaccessible through any single view alone. Live demo available at https://syndesmoscope.vercel.app/.}


\teaser{
  \centering
  \includegraphics[width=0.96\linewidth]{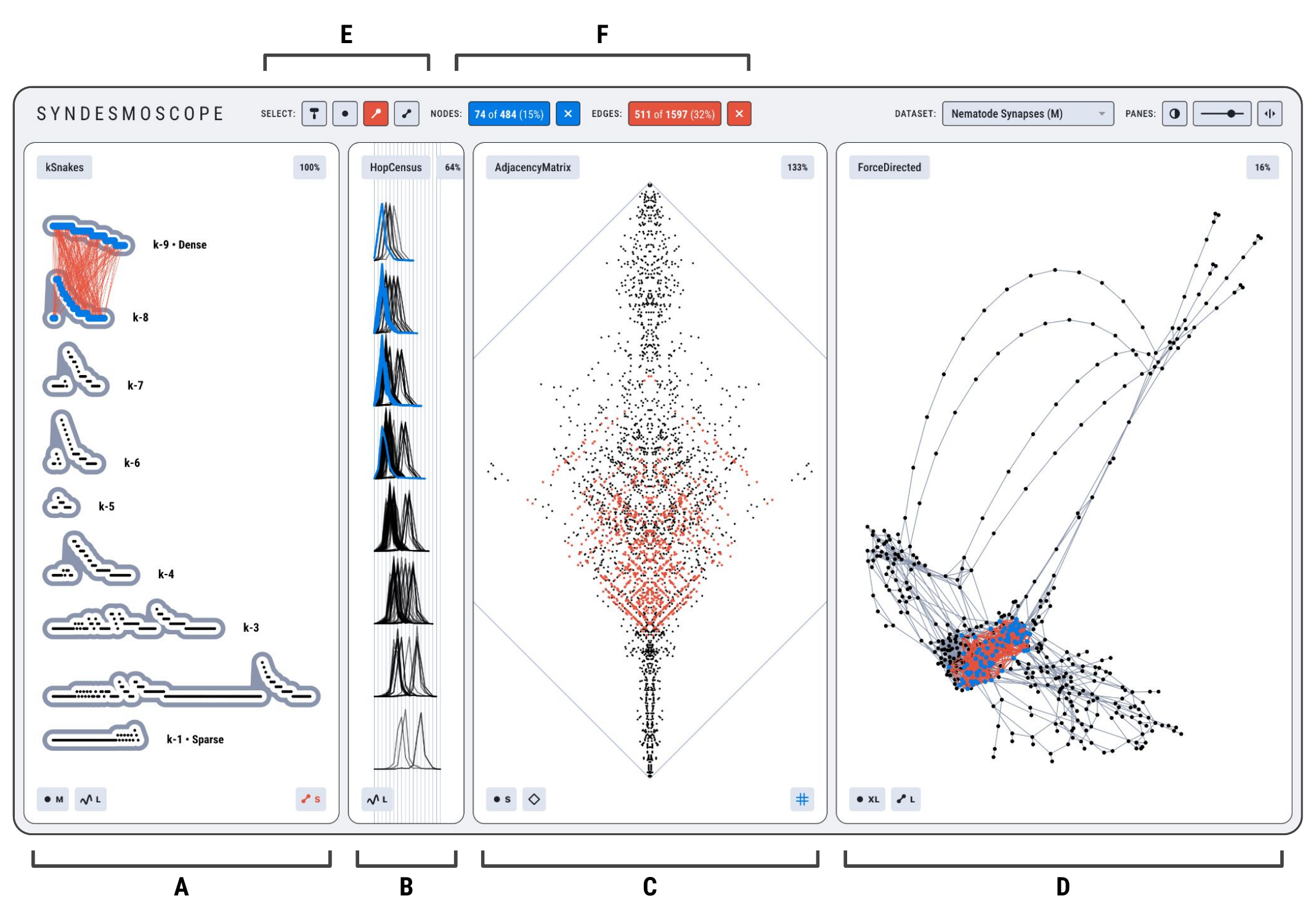}
  \caption{Syndesmoscope shows four interpretable geometric layouts computed from the same graph. \textbf{(A)} kSnakes pane, a new invariant plot based on the dense-sparse gradient; nodes from the two densest subgraphs selected at the top of the pane (blue), with hopscotching to select their edges (orange). \textbf{(B)} HopCensus pane, an invariant plot based on eccentricity~\cite{Oddo2025}. \textbf{(C)} AdjacencyMatrix pane, with a Fiedler seriation~\cite{Behrisch2016}. \textbf{(D)} ForceDirected pane, a familiar variant view. \textbf{(E)} Hopscotching controls. \textbf{(F)} Selection set counts.}
  \label{Teaser}
}

\graphicspath{{figs/}{figures/}{pictures/}{images/}{./}} 

\usepackage{tabu}                      
\usepackage{booktabs}                  
\usepackage{lipsum}                    
\usepackage{mwe}                       
\usepackage{ccicons}                   

\usepackage{mathptmx}                  

\newcommand{\SuppOne}[0]{Supp.~\S1}
\newcommand{\SuppTwo}[0]{Supp.~\S2}
\newcommand{\SuppThree}[0]{Supp.~\S3}
\newcommand{\SuppFour}[0]{Supp.~\S4}

\newcommand{\AppKSnakesTwo}[0]{App.~Fig.~11}

\begin{document}

\maketitle

\vfill\null

\section{Introduction}
\label{sec:Introduction}


\begin{figure*}[!t]
\centering
\includegraphics[width=0.96\linewidth]{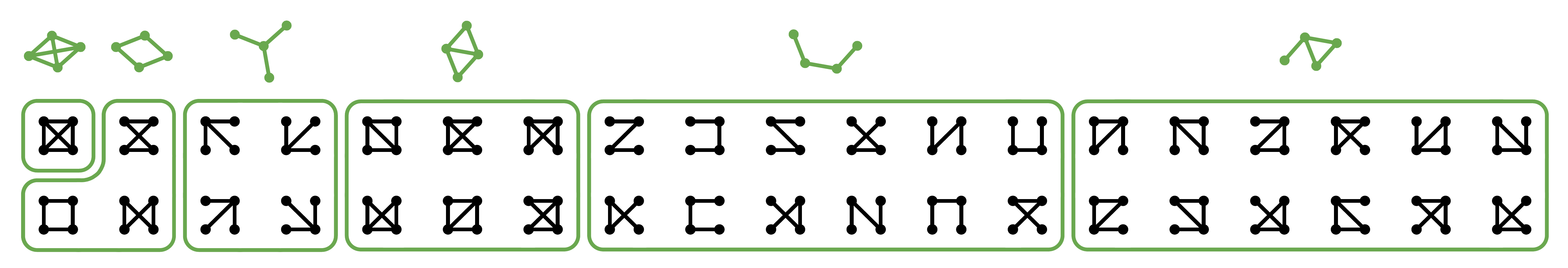}
\caption{Node-link views of the 38 labeled graphs of 4 nodes (black), with node positions fixed at the corners of a unit square to encode labels. Above the corresponding isomorphism sets (green groups) are the 6 unlabeled, non-isomorphic graphs of 4 nodes (green).}
\vspace{-10pt}
\label{IsomorphismSet}
\end{figure*}


Many network visualization idioms exist that show visual patterns that reflect structurally meaningful network patterns, each producing a distinct geometric layout tailored to specific tasks~\cite{Shu2025}. However, no single network visualization idiom affords success across all tasks~\cite{Abdelaal2023}. 

Traditional idioms such as Node-Link and Adjacency Matrix views can exhibit variability in their visual patterns, even when instantiated from the same graph topology. In contrast, \textbf{invariant plots}, as their name suggests, consistently show the same visual pattern for the same input graph~\cite{Oddo2026}. Moreover, these plots can be tightly coupled to specific topological properties of the graph, so that the geometric layouts can encapsulate targeted aspects of topological structure. Although previous work has proposed several invariant plots~\cite{Bagrow2008, Yoghourdjian2016, Oddo2025}, the visual patterns surfaced in these geometric layouts are still not well understood. We aim to address this gap in our work.

The primary contribution of this paper is Syndesmoscope, an interactive visualization system to explore a single graph through juxtaposed linked views of both invariant plots and traditional network visual encodings. This name is a neoclassical compound word built from Greek roots: 'syndesmos', which is bond or link as a noun, or "to bind" as a verb; and 'scope', which is "instrument for observing"; thus, an "instrument for observing connections". Three of the Syndesmoscope views provide distinct visual patterns that arise from the tight mapping of interesting topological properties to interpretable geometric layouts (Fig.~\ref{Teaser}-A/B/C), alongside a familiar and intuitive force-directed view (Fig.~\ref{Teaser}-D). 

Syndesmoscope features two central interaction techniques: we use the term \textbf{leapfrogging}~\cite{Renoust2015} for linked highlighting between the views to emphasize the affordances of the highly distinct visual patterns between the different views. Syndesmoscope handles selection of nodes and edges through two independent sets: one for nodes, and one for edges. With these sets the network can be traversed by \textbf{hopscotching}: that is, by extending these selection sets through topological hops from an edge to the nodes it connects, or from a node to the edges attached to it. 

Our secondary contribution is a new invariant plot we call kSnakes, which show visual patterns from the hierarchical density decomposition of graph topology, computed from the classic k-core decomposition~\cite{Batagelj2002} and the more recent complementary onion decomposition~\cite{HebertDufresne2016}. This new idiom features horizontal snake-like glyphs with a nested structure showing internal connectivity relationships, laid out along a vertical continuum from dense to sparse. 

We describe the system design and architecture, and provide an initial validation of utility through Syndesmoscope usage scenarios. We demonstrate how hopscotching and leapfrogging between the invariant plots and traditional network views reveal insights about topological phenomena that are inaccessible by any specific view alone.


\section{Background}
\label{sec:Background}

We now discuss several existing network visualization idioms that are used in Syndesmoscope, along with the graph descriptor data structures that underlie them. 

\subsection{Descriptors and Invariant Descriptors}

We can capture some of the topological information contained within graph topology into more compact data structures, known as graph descriptors. These can take many shapes and sizes, such as integers, vectors, and matrices. Some examples are the graph diameter (a scalar value), frequency counts of node degrees (a vector), and the graph's Laplacian (a square matrix). All of these examples capture only the information available within \textbf{unlabeled} graphs, where only topological placement and structural properties distinguish nodes. A single unlabeled graph represents its entire isomorphism class of labeled networks; that is, its \textbf{isomorphs}~\cite{Oddo2026}. To illustrate, with only 5 connected nodes, the possible 728 unique labeled graphs collapse into just 21 unlabeled, non-isomorphic graphs; in Fig.~\ref{IsomorphismSet} we show the scenario with 4 nodes.

Previous work proposed the term \textbf{invariant descriptors} for these unlabeled graph data structures~\cite{Oddo2025}; they are always the same for all isomorphs of a graph. Invariant plots attempt to surface visual patterns that do not change between isomorphs, so that node relabeling has no effect at all on the layout~\cite{Oddo2026}. This approach to network visualization focuses on revealing purely topological information, to avoid visual variability that arises solely from label semantics. It gives rise to very different geometric layouts than traditional network visualization idioms, where it is assumed that nodes carry persistent and typically semantically meaningful labels.

Several invariant plots have been proposed: the BMatrix Network Portrait, which aggregates generalized node degrees, or number of neighbors, into a heatmap matrix~\cite{Bagrow2019}; Graph Thumbnails, which visually maps a graph's invariant density hierarchy to concentric bubbles~\cite{Yoghourdjian2016}; and Census plots, which map invariant vectors to polylines within an absolute frame of reference~\cite{Oddo2025}. We now describe the three types of existing plots, and the descriptors that underlie them, that are used within Syndesmoscope. 

\subsection{Hop-Census Plots}

\textbf{Census-Stub vectors} are invariant descriptors that store the degree counts of the half-edges encountered through graph traversal~\cite{Oddo2025}. The vector stores high degrees as high values, and its index maps to the number of topological hops away from the source node. Collectively, the differential of degree counts within the vector encodes \textbf{degree dispersion}, a proxy for the structural uniqueness of a node with respect to the topology as a whole.

The length of the vector associated with each node may differ, because the distance to traverse the rest of the graph from each different source node varies according to topological position~\cite{Cui2022}. This Census-Stub vector length directly corresponds to the geodesic \textbf{eccentricity}~\cite{Newman2003}, where short vectors represent radius nodes at the topological \textbf{center} and long vectors represent nodes at the topological \textbf{perimeter} that are the diameter endpoints. 
 
In the Hop-Census plot (Fig.~\ref{Teaser}-B), each vector is encoded as a polyline in a shared parallel coordinate axis space. A key strength of this encoding is that the data drives the horizontal ordering of the parallel axes, in contrast to a standard parallel coordinate plot where that ordering is arbitrary. Polyline length encodes eccentricity (the number of hops in the horizontal span), and polyline shape encodes degree dispersion. Note that the Hop-Census plot only represents nodes, and does not encode edge information.

\subsection{Seriated Adjacency Matrix Views}

An Adjacency Matrix view represents a graph through a derived data table, where ordered nodes are the rows and columns, with edges represented by non-empty cells (points) between the corresponding nodes (Fig.~\ref{Teaser}-C). 
\textbf{Matrix seriation} aims to find an order for the rows and columns that creates a meaningful visual pattern of the edge-points that captures salient aspects of the topological structure. However, even carefully seriated visual patterns can be difficult to read and interpret visually~\cite{Ghoniem2004}.

The underlying data table can also be used to efficiently compute the powerful per-node Fiedler value graph descriptor, from the eigendecomposition of the Laplacian matrix. The Laplacian matrix is obtained by starting with the Adjacency Matrix and subtracting from it a diagonal matrix of node degrees. The eigendecomposition of the Laplacian results in two descriptors: the eigenvalues of the Laplacian (a vector of $N$ entries, where $N$ is the number of nodes), and the eigenvectors ($N$ in total, each with $N$ entries), with each eigenvector associated with an entry in the eigenvalues vector~\cite{Hu2013}. Among the eigenvectors, the one associated with the second smallest eigenvalue is the Fiedler eigenvector, and within the Fiedler eigenvector, each entry is the \textbf{Fiedler value} (a scalar) associated with a specific node~\cite{Jamil2023}. Practitioners commonly use a matrix seriation based on Fiedler values to surface visual patterns in the Adjacency Matrix view~\cite{Behrisch2016}.

However, although the eigenvalues of the Laplacian (a vector) is an invariant descriptor impervious to node relabeling, the Fiedler eigenvector has entries that map to graph nodes, which is sensitive to matrix seriation~\cite{Concas2022}, and therefore not fully invariant.


\section{New Invariant Plot: kSnakes}
\label{sec:kSnakes}

The kSnakes idiom is based on the intrinsic dense-sparse gradient of a graph, which is computed from the distribution of shared edges between nodes. The graph is broken down into two-level components that are distributed along a vertical axis that ranges from dense at the top to sparse at the bottom. 

\subsection{Density Invariant Descriptors}

The kSnakes idiom relies on three invariant descriptors, two proposed in previous work and one new one. The first two, while related, are different density decomposition invariant descriptors: the $k$-core~\cite{Batagelj2002} and onion~\cite{HebertDufresne2016} values. Each of these algorithms is deterministic, label-independent, and hierarchically nested, and compute a single scalar value for each node in the graph. Both algorithms compute these invariant descriptors by decomposing the graph iteratively, by removing nodes with the lowest degrees. However, the algorithms differ in how the information is stored.

The k-core algorithm~\cite{Batagelj2002} records the \textbf{shell} of value $k$ for each node in the graph, which is the threshold node degree at which it was removed from the graph during the iterative reduction. A shell (also known as a core) is simply the subgraph of all nodes with the same shell value, where high values are dense and low values are sparse. The nodes within a shell are guaranteed to have at least $k$ neighbors with nodes in the shells of value $k$ or higher. 

In contrast, the onion algorithm~\cite{HebertDufresne2016} records the exact sequential order in which each node was removed from the graph as its \textbf{onion} value. These onion values can be considered a higher-resolution version of the shell values, in that the range of onion values that belong to each shell is disjoint and monotonically increases, but the shell boundaries cannot be determined from the onion values alone. 

We also compute a new invariant descriptor, the subshell value. We partition each shell subgraph into topologically connected components, each of which is a subshell.  We then sort the subshells according to their size (number of nodes), the maximum onion value of its constituent nodes, and the cumulative onion value summed across all of its nodes. The resulting ordered index is the \textbf{subshell} value, which is also an scalar, but is associated with the higher-level construct of a subshell rather than with an individual node.


\begin{figure}[!t]
\centering
\includegraphics[width=0.99\linewidth]{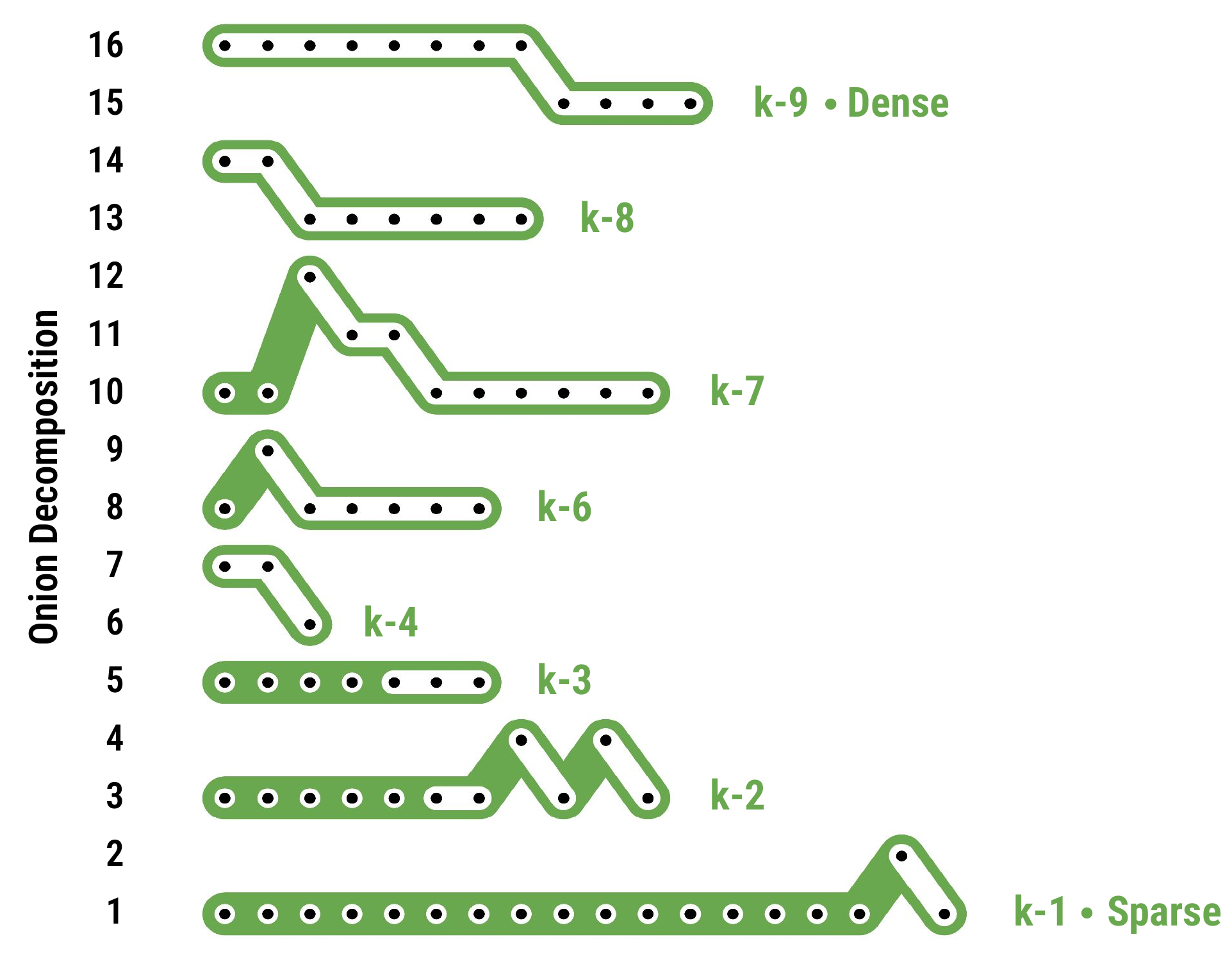}
\vspace{4pt}
\caption{The kSnakes invariant visual pattern of the 'Les Misérables' literary dataset: 77 characters (nodes) and 254 interactions (edges). Shells (green) are distributed vertically by k-core density and further broken into connected components (white) which are ordered by component size and onion density values; nodes are shown as points (black) within them. }
\vspace{-9pt}
\label{kSnakes}
\end{figure}


\subsection{Visual Idiom}

The kSnakes idiom (Fig.~\ref{kSnakes}, also Fig.~\ref{Teaser}-A) is built around the metaphor of snake-like glyphs for each shell, with topologically connected subshell components nested within them that are horizontally ordered according to both size and onion values. The snakes are vertically distributed according to their density, which is reflected in similarly nested onion value ranges that fall within the shell values. 

The idiom represents nodes as black points with equidistant spacing along a horizontal axis, which is uniform within each shell and subshell. Their vertical position depends on their onion values. These positions determine the shape of the lower layer of enclosing subshells in white, and the higher level of enclosing shells in blue (Fig.~\ref{kSnakes}). Text to the right explicitly shows the $k$ value of each shell. Fig.~\ref{kSnakes} shows the onion decomposition values only for illustration; these are not visible in Syndesmoscope.

The horizontal positions of the subshells within the shells is determined by the subshell descriptor. The ordering algorithm used to construct the descriptor yields a zigzag structure, with the largest subshells on the right, to make the subshell decomposition visually salient. 

Although Fig.~\ref{kSnakes} contains 8 shells, the $k$-core values range from $k=1$ to $k=9$. This particular topology does not have a $k$-5 shell, which illustrates an important insight: there is not a one-to-one mapping between the number of shells and the $k$-core value of the shell.

The kSnakes idiom always shows all of the nodes, and some or all of the edges can also be drawn on top of that base layer on demand. Although drawing all of the edges may lead to visual clutter, it can be instructive to show a carefully chosen subset of them. 

The kSnakes layout makes the nested structure of the graph's dense-sparse gradient visually salient, but decouples that from the underlying topological structure. For example, there is no guarantee that hierarchically adjacent shells are topologically contiguous. Furthermore, the nested subshells within a shell may fall across multiple subgraphs in the base topology.  For example, Fig.~\ref{ScenarioIntuitive} shows the extreme case of a single edge that extends all the way from the sparsest shell to the densest shell, revealing the difference between individual topological connections and this density decomposition. 


\section{Related Work}
\label{sec:RelatedWork}

We describe existing network analysis and visualization frameworks, systems that involve invariant descriptors, and linked highlighting in the context of network visualization.

\subsection{Network Data Analysis Frameworks}

Several frameworks provide rich algorithmic tools for network pattern analysis. For research, NetworkX~\cite{Hagberg2008}, which we use in the Syndesmoscope backend, is the foremost Python library for network analysis, integrating with matplotlib~\cite{Hunter2007} for visualization. With a more statistical focus, igraph~\cite{Csardi2006} and ggplot2~\cite{Wickham2016} are the R equivalent libraries. Specialized research tools like the Open Graph Drawing Framework (OGDF) offer a C++ library dedicated to graph drawing~\cite{Chimani2013}, while commercial platforms like yFiles offer developer-focused libraries that provide UI components to build network analysis applications~\cite{yWorks}. All of these frameworks require programming expertise to explore and visualize graph topology, while Syndesmoscope does not. 

\subsection{General Network Visualization Systems}\label{Sec:GeneralNetworkVis}

Accessible network visualization systems such as Cytoscape~\cite{Shannon2003} and Gephi~\cite{Bastian2009} are user-friendly and support broad plugin ecosystems that focus on bioinformatics and social science, respectively~\cite{Pavlopoulos2017}. Cosmograph ~\cite{Cosmograph} is similar to Gephi, but GPU-accelerated to handle large-scale networks. Pajek~\cite{Batagelj1998} also handles large datasets, though without an approachable interface for nontechnical users. All of these systems overly rely on the Node-Link idiom. In contrast, the powerful Tulip~\cite{Auber2018} visualization system involves multiple views with linked highlighting. However, none of these systems include invariant plots in their views.

\subsection{Invariant Descriptor Visualizations}

The two closest idioms to our  kSnakes plots occupy two extreme points along the spectrum from cluttered to aggregated. LaNet-vi~\cite{AlvarezHamelin2005} draws all nodes as points laid out concentric shells based on $k$-core decomposition, but with high levels of clutter from overplotting. Graph Thumbnails~\cite{Yoghourdjian2016} is a highly aggregated representation that draws only the shells as nested bubbles, with no visible encoding at the node level. In contrast, kSnakes occupies a more effective middle ground, capturing the graph's intrinsic dense-sparse gradient levels as explicitly separated structures while still showing node-level information.

The GraphPrism system~\cite{Kairam2012} is the closest to Syndesmoscope in concept and design, juxtaposing a traditional Node-Link view with multiple faceted instantiations of the same invariant plot, a version of the BMatrix~\cite{Bagrow2008}. Syndesmoscope features the HopCensus invariant plot~\cite{Oddo2025} which is a de-aggregated generalization of the BMatrix that provides full node-level information. It also supports two additional idioms, our new kSnakes plot and an Adjacency Matrix view with a specialized seriation, for more powerful exploration support. Also, although GraphPrism is described as a D3 application, the authors did not make a working implementation available, whereas Syndesmoscope is released as open source.

\subsection{Linked Highlighting in Network Visualization}

Linked highlighting is a foundational interaction idiom within the Coordinated Multiple Views paradigm~\cite{Roberts2007}, allowing users to observe the same data through different visual encodings simultaneously. In network visualization, MatrixExplorer~\cite{Henry2006} is a dual-representation system that links the traditional Node-Link and Adjacency Matrix views. Building on complementary views, Renoust et al~\cite{Renoust2015} show how linked highlighting can manage the complexity of multiplex networks through the leapfrogging interaction idiom, a specialized form of linked highlighting that allows users to seamlessly alternate their exploratory focus between different specialized Node-Link views. Syndesmoscope generalizes the idea of leapfrogging from traditional idioms to novel invariant plots.


\section{System Design}

We provide an overview of Syndesmoscope design, its support for network tasks that do not depend on the existence of labels~\cite{Oddo2026}, and panes with views that tightly map interesting topological properties to interpretable geometric layouts. In addition, we describe the user interface, view coordination, and system architecture.

\subsection{Tasks and Motivation}

A recent taxonomy of tasks for invariant plots~\cite{Oddo2026} frames tasks according to five scopes: the  Constituent level of node and edge concerns, the Subgraph exploration level, the Single graph level, the Pairwise comparison between graphs level, and the Multiple graph level. Syndesmoscope focuses on the first three levels, which we collectively refer to here as the mesoscale scope. We define \textbf{mesoscale phenomena} as encompassing both known and yet-undescribed Subgraph patterns that lie between the Constituent scope of nodes and edges and the Single scope of the graph as a whole~\cite{Oddo2026}. Mesoscale phenomena range from easy-to-define, low-level structures such as stars, chains, triangles, and cliques; to more fuzzy Subgraph structures such as near-trees and clusters~\cite{Girvan2002}; to more sophisticated Single graph classifications such as recurring mesoscale patterns~\cite{Lee2014} or core-periphery hierarchy~\cite{Borgatti2000} (which we call dense-sparsity gradient in this paper). 

This taxonomy also distinguishes between labels-required tasks that cannot be carried out with invariant plots based on unlabeled graph data abstractions, and labels-not-required tasks that can be supported by both invariant and traditional plots~\cite{Oddo2026}. For example, "Find node A and node B and trace the path between them" is a labels-required task: it begins by identifying nodes with semantic labels A and B, which then constrains the path-tracing step if the labels are not available. In contrast, labels-not-required tasks include "Compare two or more structurally similar nodes" (see Fig.~\ref{ScenarioSubgraphs}), "Where are the dense nodes in this graph" (see Figs.~\ref{Teaser}, \ref{kSnakes}, \ref{ScenarioIntuitive}, and \ref{ScenarioDensity}), and "Locate and characterize bridge edges" (see Fig.~\ref{ScenarioBridges}).

Syndesmoscope is designed to support labels-not-required tasks at the Constituent, Subgraph, and Single graph scopes, including complex cases that are not well supported by prior systems that lack invariant plots.

\subsection{System Overview}

The goal of the Syndesmoscope system is to help its users better interpret, describe, and characterize mesoscale phenomena in graph topology. We do so with four juxtaposed panes, each dedicated to a network visualization idiom that show a different visual pattern. The kSnakes, HopCensus, and AdjacencyMatrix panes feature a tight mapping of interesting topological properties to interpretable geometric layouts. The ForceDirected pane supports overall sensemaking by providing a more familiar representation linked to the potentially unfamiliar visual patterns in the other three views.

We instantiate the selection interaction to be independent between node sets and edge sets. The interplay between these two sets -- within or across panes -- affords sophisticated linked highlighting from which to compose interactive exploration of mesoscale phenomena.

More specifically, hop-based traversal can systematically extend the selected sets through an interaction we call \textbf{hopscotching}, which is gradually extending selected data from nodes to their connecting edges, or vice versa from a set of edges to all of the nodes that they connect. Hopscotching allows us to produce visual patterns in the display by traversing network patterns in the underlying graph data structure.

This ability to quickly explore topological connections, hop by hop, is particularly important because of what it reveals in the visual patterns visible across the different views. We emphasize this power by using the term \textbf{leapfrogging}~\cite{Renoust2015} to mean linked highlighting between different views with distinct visual patterns, so selection of a contiguous area of geometric layout in one view is likely to illustrate different patterns in the other views that reveal different aspects of the underlying network patterns.

We call out these two approaches with specific names to emphasize how in conjunction with each other, they afford the ability to explore novel visual patterns revealing underlying network patterns, within and across the views.

\subsection{Visual Patterns from Interpretable Axes}

We tightly map topological properties from different corners of graph theory to interpretable geometric layouts: Sparsity Axis, Eccentricity Axis, and Fiedler Axis. We contrast these axes to the force-directed layouts mapped onto the Force-Directed Plane. The visual patterns are made of marks, which are specific to each pane: node-points as a base layer with optional edge-line overlays in kSnakes, node polylines in HopCensus, edge-points as a base layer with optional node gridlines in AdjacencyMatrix, and both node-points and edge-lines in ForceDirected.

\begin{figure}[!t]
\centering
\includegraphics[width=0.99\linewidth]{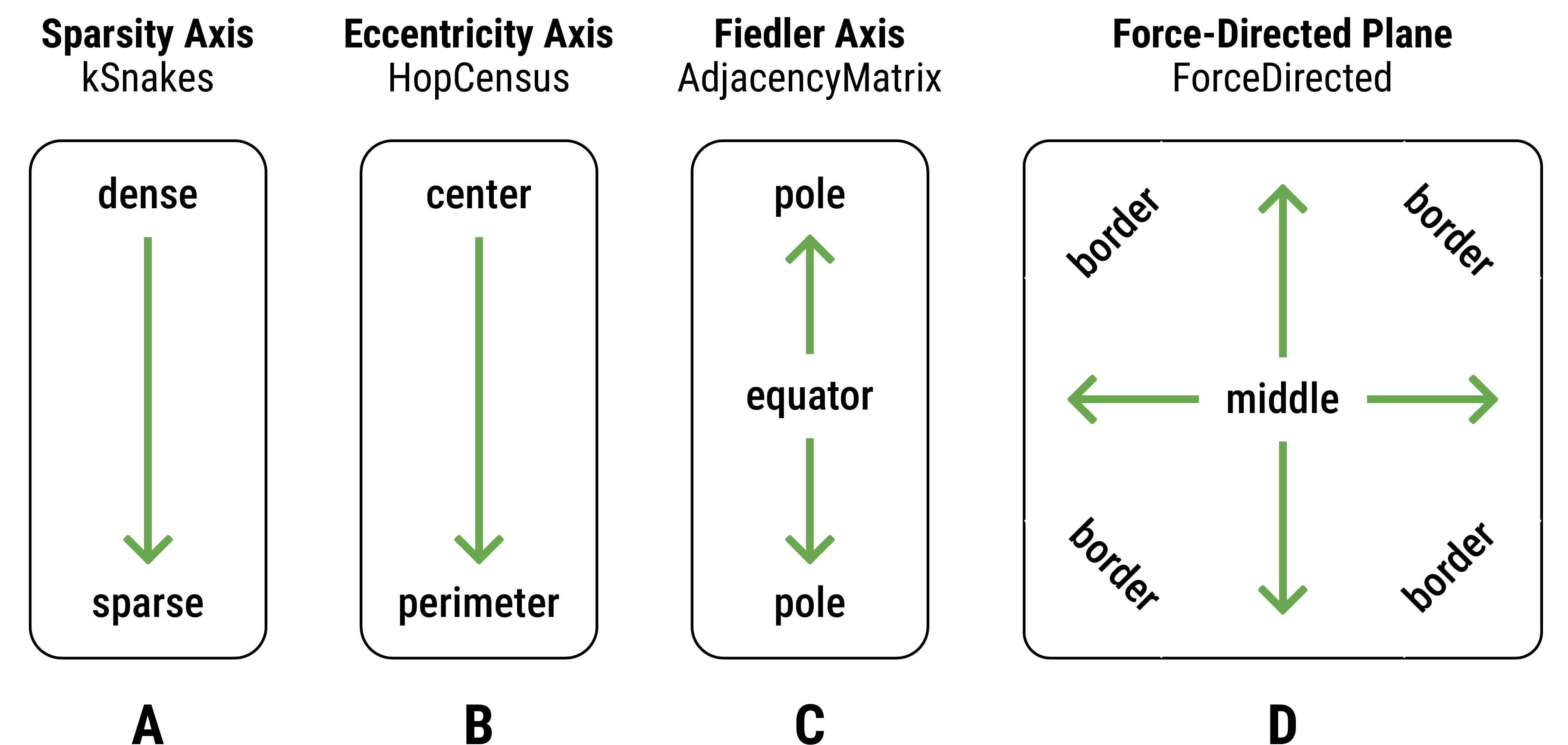}
\caption{Interpretable axes in Syndesmoscope panes. \textbf{(A)} The Sparsity Axis in kSnakes increases from top to bottom. \textbf{(B)} The Eccentricity Axis in HopCensus increases from top to bottom. \textbf{(C)} The Fiedler Axis in the AdjacencyMatrix pane is divergent, from the equator to the poles. \textbf{(D)} The Force-Directed Plane has two divergent spatial axes.}
\vspace{-10pt}
\label{Axes}
\end{figure}

\textbf{Sparsity Axis.} In the kSnakes pane, the vertical placement of node-points is determined by the graph's intrinsic dense-sparse gradient, ranging from \textbf{dense} at the top to \textbf{sparse} at the bottom (Fig.~\ref{Axes}-A). The layout groups node-point marks into glyphs representing shells of equal density, allowing easy selection of nodes based on their sparsity characteristics.

The visual pattern further supports more granular selection through the visual distinction of subshells of connected components enclosed within the shells. For this axis we use the terms dense and sparse to describe its extremes, rather than the traditional terms core and periphery~\cite{Borgatti2000}, as the latter terms clash with concepts from the other algorithmic lenses in this paper.

\textbf{Eccentricity Axis.} In the HopCensus pane, the layout collects node polylines into separate bins according to polyline length, where the shape of the polylines within each bin encodes the degree dispersion. The layout vertically distributes bins by eccentricity value, from the topological \textbf{center} at the top to the topological \textbf{perimeter} at the bottom (Fig.~\ref{Axes}-B). The vertical separation of HopCensus polylines by eccentricity into bins is a further de-aggregation from the original BMatrix~\cite{Bagrow2008, Bagrow2019}.

The valence of the Eccentricity Axis matches that of the Sparsity Axis, to reflect the semantics of geodesic eccentricity: nodes with highest eccentricity are diameter endpoints on the perimeter, with low degrees, therefore sparse nodes. Thus, we visually map maximal eccentricity and maximal sparsity to the bottom of their respective views, to ensure horizontal visual continuity between the kSnakes and HopCensus panes. The resulting HopCensus visual pattern affords the easy selection of equal-eccentricity polylines, which belong to the same vertical bin. Within a bin, the selection of individual polylines with the same local shape, or degree dispersion, yields nodes of the same degree count at that depth of traversal.

\textbf{Fiedler Axis.} The AdjacencyMatrix pane deals with edges, in contrast to the Density Axis and Eccentricity Axis that are primarily based on nodes. We present the AdjacencyMatrix idiom in two ways that differ from the standard representations.

First, we orient the matrix so that the Fiedler axis is vertical rather than on the diagonal. There is a \textbf{pole above} and a maximally opposite \textbf{pole below}, with an \textbf{equator} band in the middle, equidistant to both poles (Fig.~\ref{Axes}-C). The resulting visual pattern affords the easy selection of edge-points at the same vertical height through rectangular selection. If the orientation were the more traditional square, it would be more difficult to select all of the points at the same relative distance to the poles. 

Second, Fiedler values are associated with nodes and come from eigenvectors, which come from the eigendecomposition of the Laplacian matrix. We leverage Census-Stub vectors to pre-seriate the Laplacian before the spectral Fiedler value computation. After isomorphic randomization, for some topologies a Census-Stub pre-seriation makes the visual patterns close to invariant (see~\SuppOne~for more details).

\textbf{Force-Directed Plane.} The ForceDirected pane shows a Node-Link representation of the graph, obtained via standard force-directed layout optimization. Such representations are familiar but highly visually variable. In graphs with simple structure, we expect high-degree central nodes to be near the \textbf{middle} of the visual plane; while low-degree perimeter nodes are closer to the \textbf{borders} of the plane (Fig.~\ref{Axes}-D).


\subsection{User Interface and Interaction}

Syndesmoscope prioritizes the visual patterns within panes, so the majority of screen space is dedicated to the juxtaposed panes (Fig.~\ref{Teaser}-A/B/C/D). Panes resize arbitrarily, allowing users to allocate screen space according to their analytical focus. All panes support scroll-wheel zooming and click-drag panning navigation, with a zoom reset toggle button on the top right corner. On the bottom left corner are visual encoding resize buttons, which cycle through pre-determined mark sizes (XS, S, M, L, XL).

Above the panes is the control strip, which includes selection controls, a dataset dropdown menu, and general interface options: a dark theme toggle, a brightness slider for non-selected monochrome patterns, and an equal-width panes toggle (shown in Fig.~\ref{ScenarioIntuitive}). The four selection controls (Fig.~\ref{Teaser}-E) include a brush toggle that enables click-drag rectangular selection; while active, panning is disabled but zooming remains available. The three hopscotching buttons, for node, stub, and edge sets, respectively, expand the corresponding sets when un-selected but available in the underlying data structure. The node set is all the un-selected nodes that connect to selected edges, the stub set is all the edges that connect to any selected nodes, and the edge set is all the edges that connect between selected nodes.

The interface reports node and edge selections as percentages of total node and edge counts, with a reset button to clear each selection set (Fig.~\ref{Teaser}-F).

A label in the top left corner identifies the pane type, which is also a button that cycles through pane types; repetitions are allowed, so there can be multiple panes showing the different corners of the same visual pattern. The kSnakes and AdjacencyMatrix panes have pane-specific buttons on the bottom right corner. In the kSnakes pane, users can show selected edges as an overlay layer, to explore how edges span the dense-sparse gradient; and in the AdjacencyMatrix pane, users can toggle the node gridlines into view. In addition, in AdjacencyMatrix the diamond button can toggle the visual pattern between square and diamond orientations.


\begin{figure*}[!t]
\centering
\includegraphics[width=0.99\linewidth]{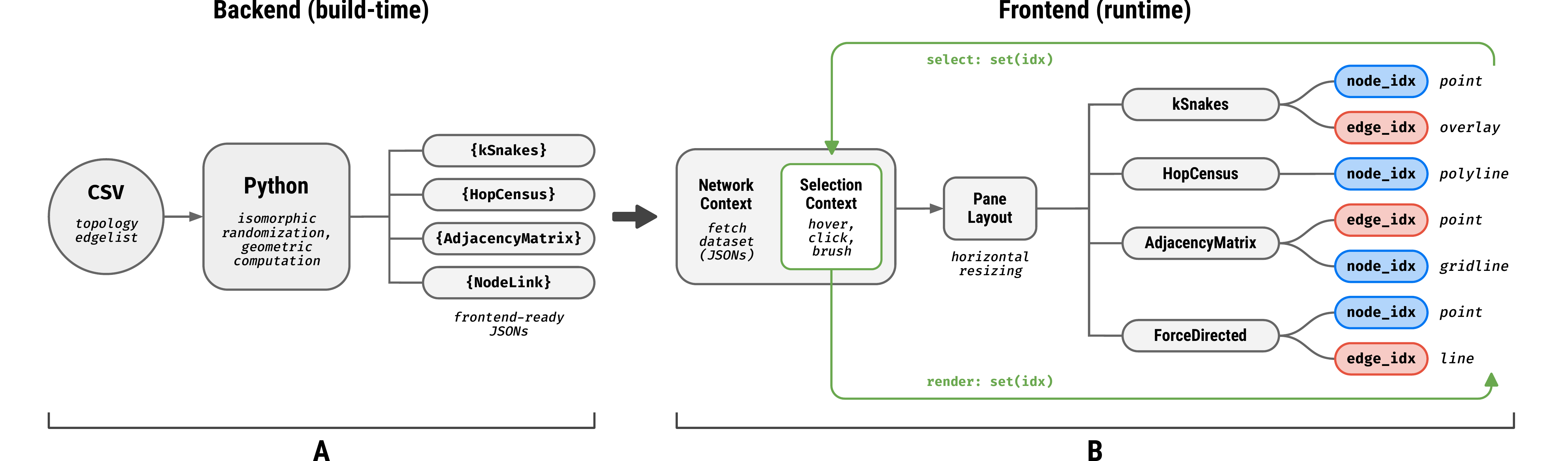}
\vspace{4pt}
\caption{The Syndesmoscope system architecture is divided into two parts, backend and frontend. \textbf{(A)} The backend is a sequential Python pipeline that computes JSONs from source data (a CSV edgelist). The JSONs contain D3-ready geometric layout information for the respective panes, all relationally linked through a shared index set (node\_idx, edge\_idx). \textbf{(B)} For a given network dataset, the interactive React frontend loads the respective JSONs. At runtime, React handles changes to index set selection; and supports hovering, clicking, brushing, and pane resizing interactions. Through view coordination (green arrows), D3 user interaction in one pane writes index set changes, which React broadcasts back to all panes, then D3 re-renders. In parallel, D3 handles the force simulation in the ForceDirected pane at runtime.}
\vspace{-8pt}
\label{Architecture}
\end{figure*}


\subsection{Architecture}

The Syndesmoscope architecture is separated into a backend that runs computationally intensive preprocessing operations on the datasets, and a frontend for rendering and interaction (Fig.~\ref{Architecture}).

The backend (Fig.~\ref{Architecture}-A) is a linear Python pipeline that runs once, at build-time, on the developer's machine. It loads a simple CSV edgelist and leverages the NumPy~\cite{Harris2020} and NetworkX~\cite{Hagberg2008} libraries for linear algebra and graph-theoretic algorithms, respectively. The pipeline stores and exports the computations as D3-ready JSONs with a shared relational index set: \textbf{node\_idx} behind the node marks, and \textbf{edge\_idx} for the edge marks. Each JSON is dedicated to a geometric layout for a specific view (see~\SuppTwo~for data structure details).

The frontend (Fig.~\ref{Architecture}-B) is a React (v18.3) and D3 (v7.9) web application. For a given dataset, React fetches the suite of JSONs in parallel, and stores them in the NetworkContext state container. With data loaded, the SelectionContext state handler updates the selection index set. The PaneLayout container manages pane types, the arbitrary horizontal resizing of the four pane slots in landscape mode, and the single pane slot in portrait mode. At runtime, React handles changes to the selection index set from the hover, click, and brush interactions. In parallel, D3 handles the force simulation in the ForceDirected pane, which features a node-point draggable interaction~\cite{Bostock2011} to manually manipulate the layout.

Although Syndesmoscope is designed with strong focus on invariant plots that apply to unlabeled graphs, actual storage in computer memory necessarily instantiates explicit labels, as does linked highlighting between views. The system thus does use the index labels as described above, for both nodes and edges, but these are temporary labels without semantic meaning. 

Moreover, we consider the issue that the implicit position of entries in a data structure may carry semantic information that is extraneous to graph topology. For instance, nodes stored in order of data collection may cluster visually in ways that reflect the collection process rather than mesoscale phenomena. To mitigate this, we implement \textbf{isomorphic randomization}, which generates a structurally equivalent version of the input graph with randomized labels and also storage configuration. This reduces the likelihood that visual patterns in our invariant plots arise from ordering artifacts rather than underlying graph topology.

After randomization, we seriate the nodes according to Census-Stub vectors, in order to have a repeatable baseline visual pattern. This seriation is most crucial for the AdjacencyMatrix view. It may also be visible in the kSnakes view because it determines the left to right order of nodes within a shell in some cases, that is, when the onion values do not differentiate between them (See~\SuppOne~for more details).

\subsection{View Coordination}

Nodes and edges have independently selectable sets. When unselected, their visual encodings are monochrome grey, and the colors change to blue for nodes (\texttt{\#007be6}) and orange for edges (\texttt{\#e6533f}) once selected. These colors work well against both a light and dark background; we found them with the Color Buddy tool~\cite{McNutt2025}.

The objects are rendered from back to front in the following order to ensure that selected items remain visible: in back are the unselected monochrome edges, followed by unselected monochrome nodes, selected orange edges, and selected blue nodes at the front. The system further emphasizes selected visual encodings by increasing size, saturation, and luminance of the visual marks.

Any pane can either initiate or receive selections through a shared “one pane writes, all panes read” model based on the relational index set. As shown in the selection state cycle in Fig.~\ref{Axes}-B, user interactions with D3 visual encodings in one pane update this shared selection state, which React propagates to all panes so D3 can re-render the corresponding visual encodings consistently across views.

\subsection{Performance}

Syndesmoscope delivers fluid frontend performance of rendering datasets of up to ~4,000 nodes and ~8,000 edges without any noticeable lag on a MacBook Pro M3 laptop. The React architecture also supports responsive rendering so that Syndesmoscope will function on displays of any size, including smartphones. Responsiveness is further optimized through a dedicated portrait mode, handled by the PaneLayout container (Fig.~\ref{Axes}-B), that displays a single, non-resizable pane; it is triggered when the height of the browser window is larger than the width (see~\SuppThree~for example). Datasets of up to ~5000 nodes and ~50,000 edges are at the upper limit of performance, where the rendering has noticeable delays, with a few seconds per frame. 



\begin{figure*}[!t]
\centering
\includegraphics[width=0.95\linewidth]{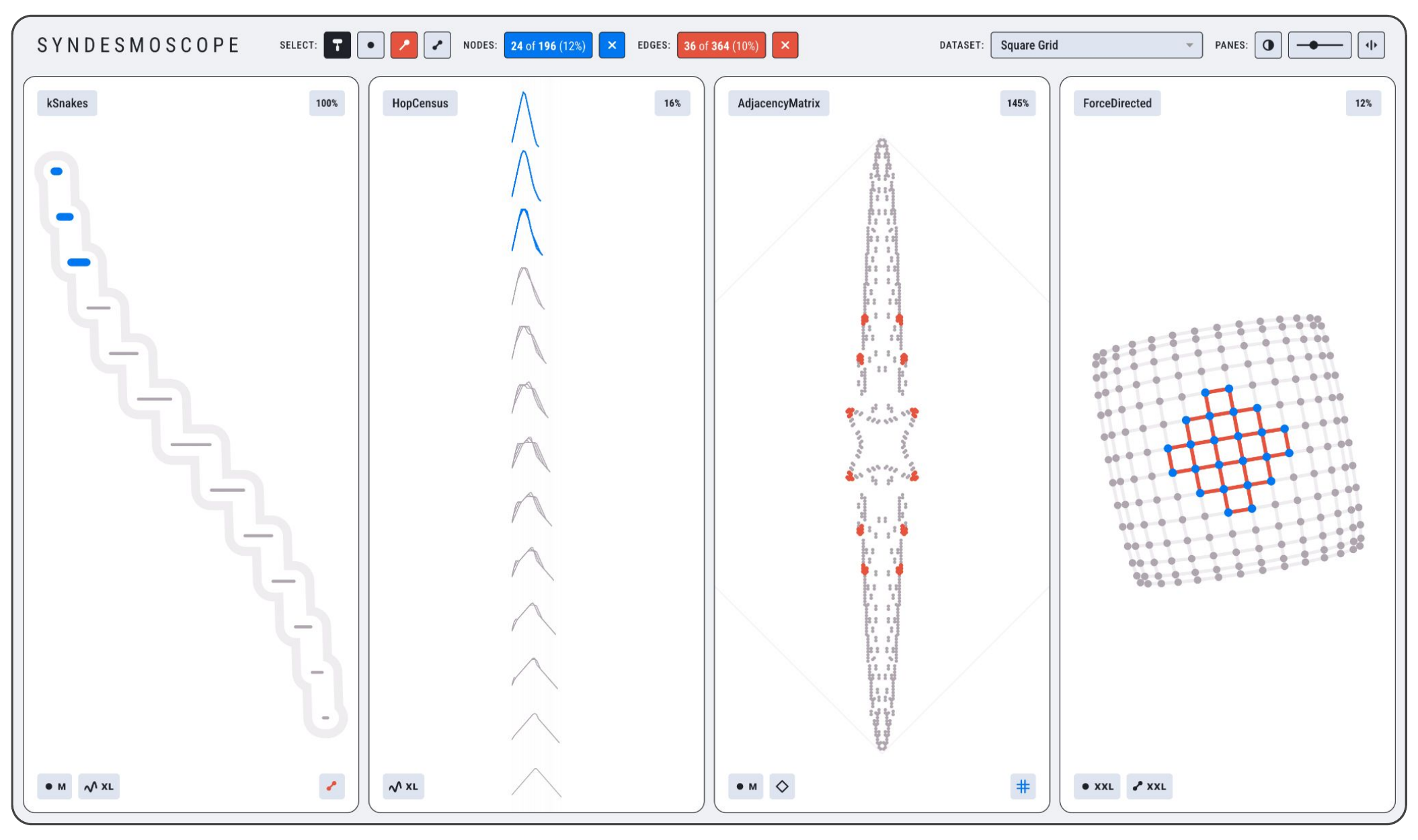}
\caption{The 'Square Grid' example dataset: 196 nodes and 364 edges, a regular lattice of 14 nodes per side. We select the densest shell nodes in kSnakes (top of the pane), which propagate to the central node polylines in HopCensus (top of the pane). We hopscotch their shared edges, which are evenly spread around the equator of AdjacencyMatrix (middle of the pane), and in the visual middle of the ForceDirected pane.}
\label{ScenarioIntuitive}
\end{figure*}



\begin{figure*}[!t]
\centering
\includegraphics[width=0.95\linewidth]{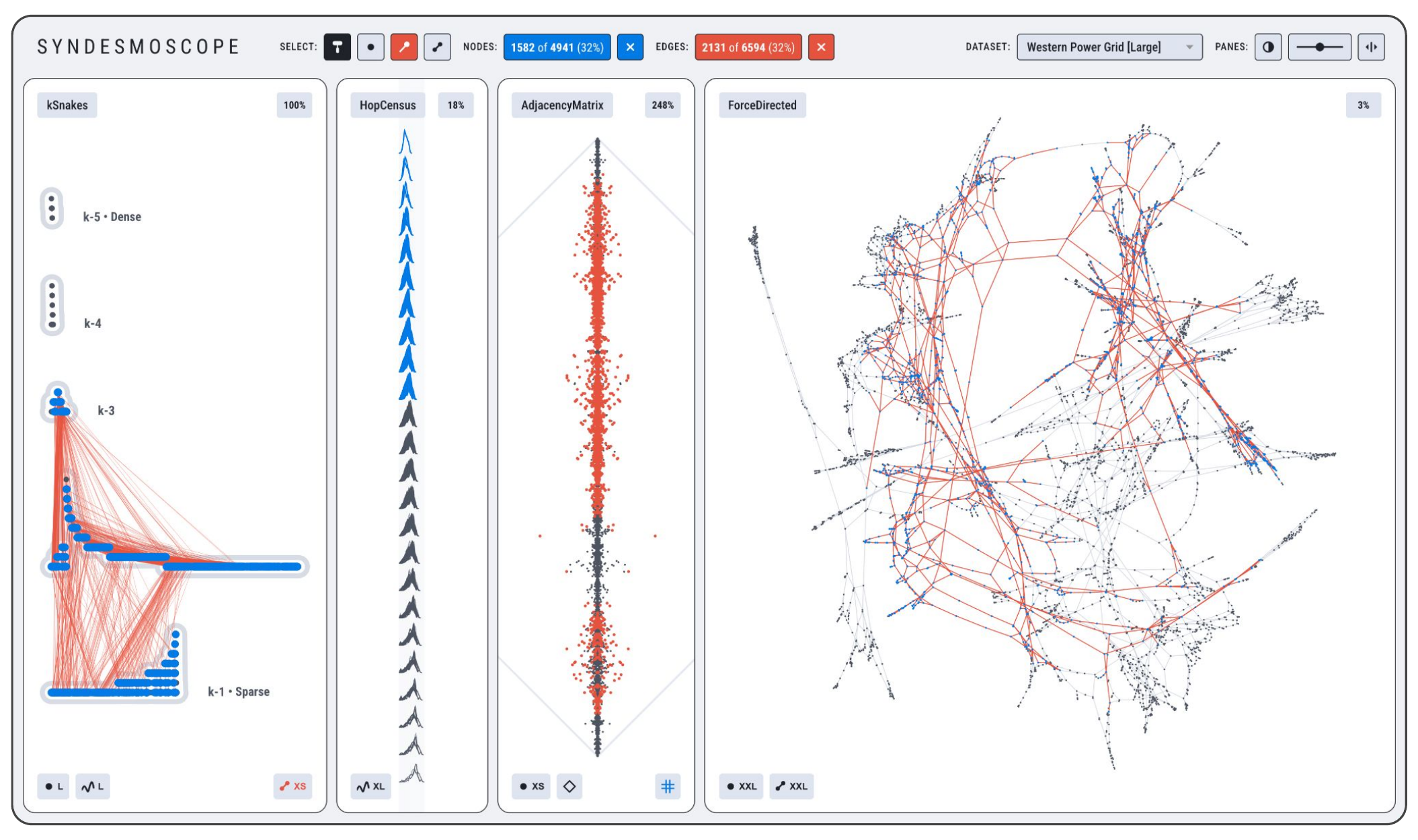}
\caption{The 'Western Power Grid' infrastructure dataset: 4941 substations (nodes) and 6594 transmission lines (edges), exhibits a counterintuitive topology with respect to sparsity and eccentricity. In the HopCensus pane, the low eccentricity polylines (top half of the pane) are central nodes that unexpectedly map to sparse nodes in kSnakes (bottom half of the pane). After further inspection, the densest shell nodes in kSnakes (top half of the pane, not selected) map to nodes far away from the main topology, we annotate their placement in the ForceDirected pane (green arrow).}
\vspace{-10pt}
\label{ScenarioDensity}
\end{figure*}



\begin{figure*}[!t]
\centering
\includegraphics[width=0.95\linewidth]{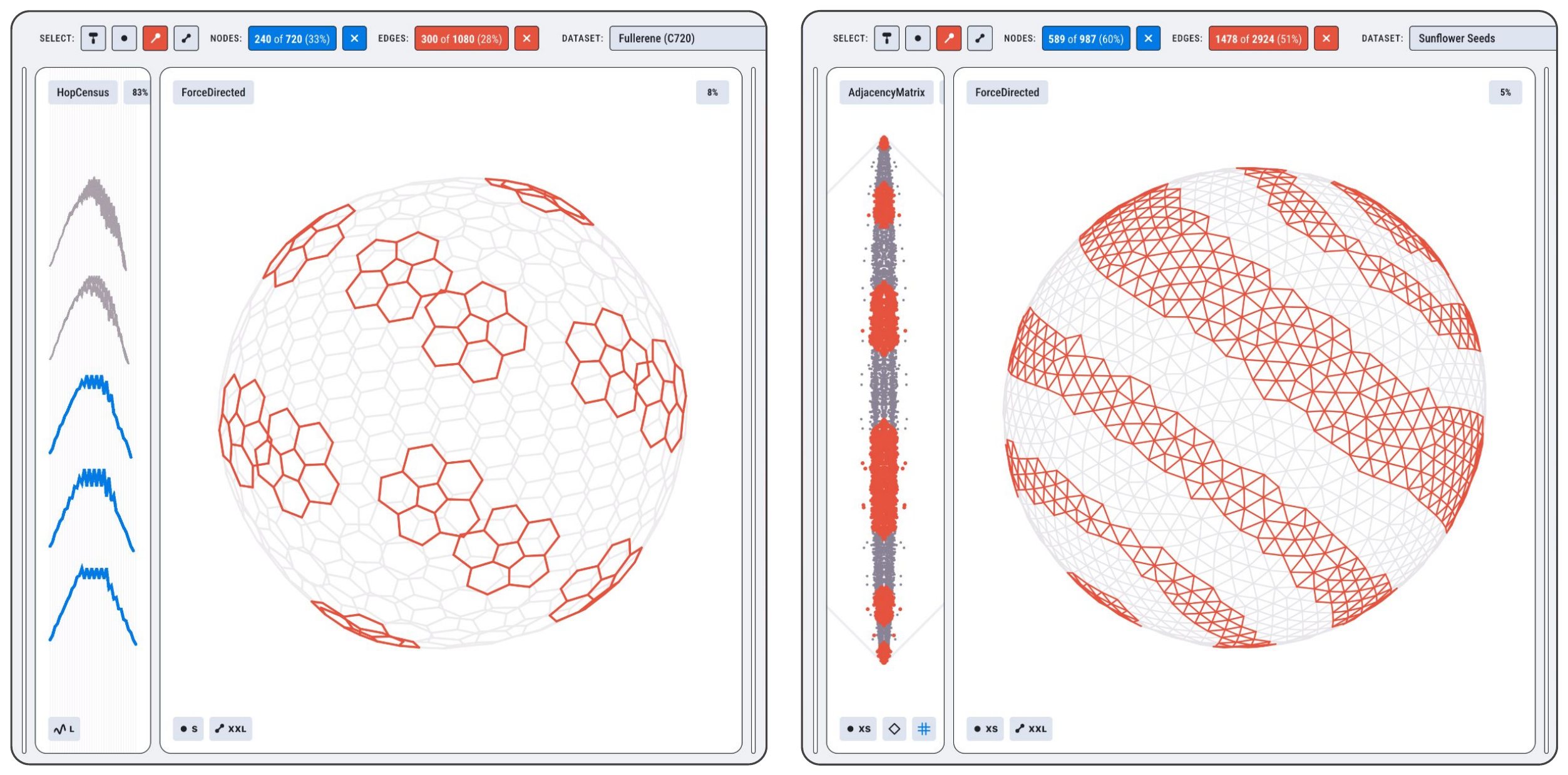}
\caption{Two separate instances of Syndesmoscope loaded with geometric datasets, each instance with only 2 of the 4 panes visible. On the left, 'Fullerene (C720)' is a spherical mesh of 720 carbon atoms (nodes) and 1080 bonds (edges), with recurring pentagonal floret motifs selected. On the right, 'Sunflower Seeds' is a planar mesh of 987 seeds (nodes) and 2924 neighbor connections (edges). }
\label{ScenarioSubgraphs}
\end{figure*}



\begin{figure*}[!t]
\centering
\includegraphics[width=0.95\linewidth]{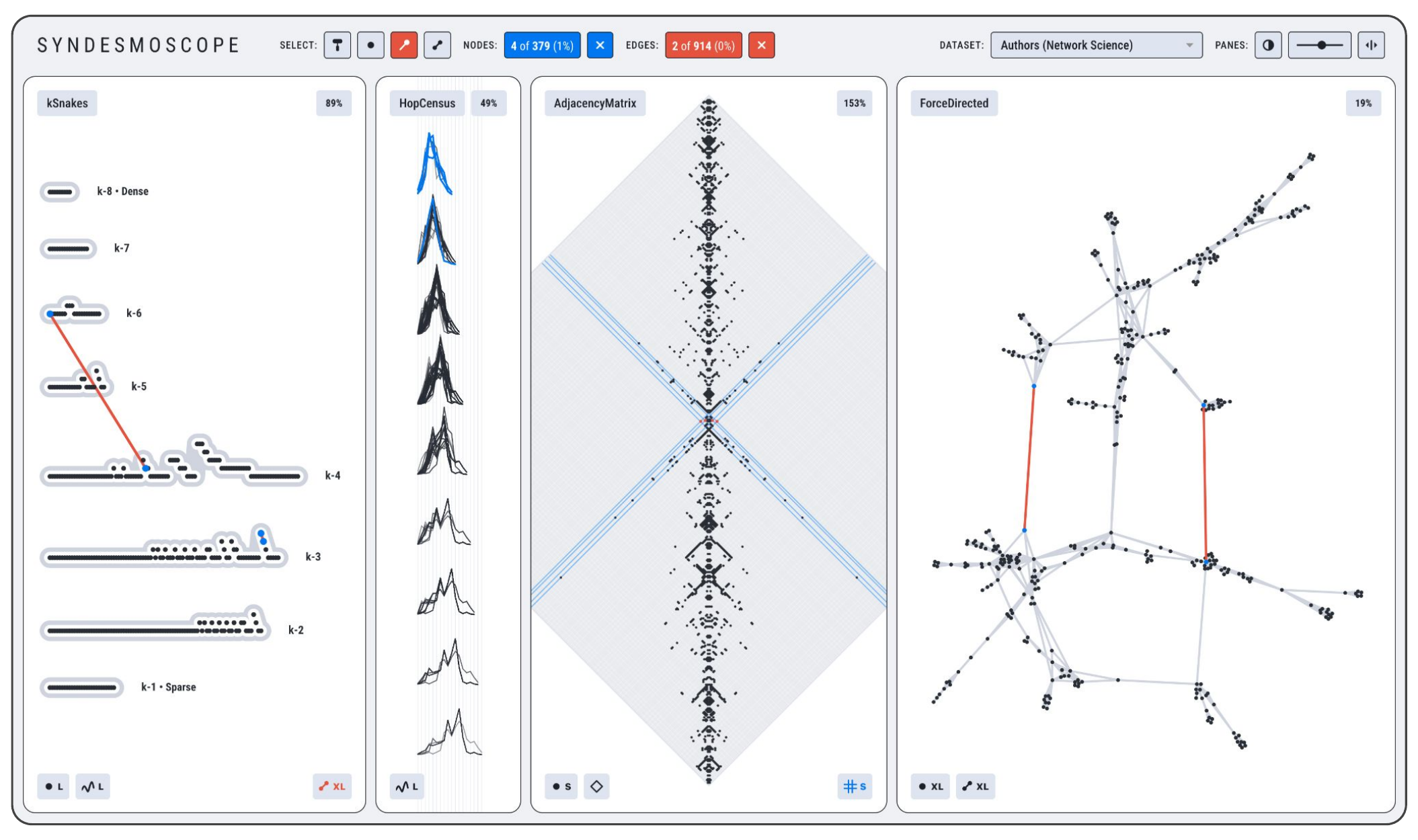}
\caption{The 'Authors (Network Science)' social network dataset: 379 authors (nodes) and 914 collaborations (edges). In the AdjacencyMatrix, highlighting edge-points in the equator reveal two bridge edges, and a complex bridge-like structure.}
\label{ScenarioBridges}
\end{figure*}


\section{Usage Scenarios}

We showcase the power of leapfrogging and hopscotching across the visual patterns and their underlying network patterns. 

\subsection{Dataset Corpus}

Syndesmoscope features a corpus of 72 real-world and synthetic networks, found in the dropdown menu of the control strip. We collected this corpus from previous work~\cite{Oddo2025} and online repositories~\cite{Netzschleuder}, and categorize into 10 domains (counts in parentheses): literary (6), ecological (4), social (12), infrastructure (8), neurological (8), biological (12), and technological (4) for the empirical networks; and simple (5), geometric (9), and synthetic (4) for the generated networks. In~\SuppFour~we share dataset corpus metadata.

In the next section we showcase a representative sample of datasets. In the Appendix we provide more usage scenarios across more datasets.

\subsection{Intuitive Example}

In Fig.~\ref{ScenarioIntuitive}, we showcase the visual patterns that align with the expected interpretation of the pane axes. We select nodes from the 3 densest shells in kSnakes (top of the pane) for the given grid graph. This selection propagates to HopCensus, where the polylines with lowest eccentricities, or the central nodes that are far away from the perimeter (top of the pane), get highlighted. By hopscotching their shared edges, in AdjacencyMatrix these edges are revealed to cluster equidistantly around the equator (middle of the pane). Finally, the ForceDirected panes shows that these 24 densest nodes are contained in the middle  of the view.

Fig.~\ref{Teaser} also aligns with the standard axes interpretation. In this example, the 'Nematode Synapses (M)' neurological dataset represents the nervous system of the male C. elegans roundworm: 484 cells (nodes) and 1597 synapses (edges). The selection of the nodes from the 2 densest shells in kSnakes propagates to low eccentricity polylines in HopCensus, and to edge-points around the equator in AdjacencyMatrix. The ForceDirected pane shows these 74 nodes selected as a dense subgraph with a central topological placement.

\subsection{Counterintuitive Sparsity}

Next we consider the Western Powergrid graph, shown in Fig.~\ref{ScenarioDensity}.  We select the ten bins at the top of the HopCensus pane which are the low-eccentricity center polylines, but leapfrogging to the kSnakes pane shows that the sparse shells on the bottom are highlighted, not the dense ones at the top. This situation is surprising, since frequently the vertical axes of the kSnakes and HopCensus panes show similiar visual patterns, by design (Fig.~\ref{Axes}-A/B). In many examples in this paper, including Fig.~\ref{Teaser} and Fig.~\ref{ScenarioIntuitive}, 
nodes with the shortest polylines in HopCensus are also nodes in the graph's topological center (lowest geodesic eccentricity), and therefore naturally reside in the kSnakes densest shell. When highlighted, these dense nodes with short polylines appear in the top regions of their respective panes, and also in the middle visual pattern regions of AdjacencyMatrix and ForceDirected panes. In~\AppKSnakesTwo, we show the result of further exploration of this dataset. 



\subsection{Recurring Subgraphs}

The 'Fullerene (C720)' graph has hexagonal and pentagonal faces. In Fig.~\ref{ScenarioSubgraphs} (left side), we select the longest polylines in HopCensus in the bottom half of the pane, or the 3rd, 4th, and 5th bins. This selection isolates the nodes with the highest geodesic eccentricity, then by hopscotching, their shared edges reveal pentagonal ``floret'' motifs (pentagon surrounded by hexagons) in the molecule.

In the 'Sunflower Seeds' graph shown in Fig.~\ref{ScenarioSubgraphs} (right side), selection of arbitrary bands in the vertical span of the AdjacencyMatrix pane reveals that edge-points that are proximal in Fiedler bisection distance are also topological proximal, and reveal a poleward pattern in the ForceDirected view that show as a wave-like gradient across the mesh.

\subsection{Bridge Characterization}
Here we consider a co-authorship network. Finding bridge edges is a well-known network analysis task, previously explored with the same network in Fig.~\ref{ScenarioBridges} with the GraphPrism system~\cite{Kairam2012}. In the AdjacencyMatrix pane, edge-points in the equator region of the visual pattern are equidistant from the top and bottom bisection poles. Highlighting the edge-points in this equator reveals the topology's two bridge edges that span roughly equal-size subgraphs, made salient by manually adjusting the visual pattern in the ForceDirected pane.

In HopCensus, the nodes belonging to these bridges exhibit extremely low eccentricity (top polyline), so positionally central nodes. In kSnakes these nodes are shown to belong to the medium density shells, with the left bridge edge in the ForceDirected pane spanning nodes from the k-4 to the k-9 shells, while the bridge edge on the right is entirely contained within the sparser k-3 shell. 

On closer inspection, there is a third bridge-like structure between both highlighted edges, seen in the middle of the ForceDirected pane. To the rest of the topology, this structure behaves like a third bridge, where cutting these three bridges would separate the topology into two roughly equal-size components. However, unlike the two highlighted bridges, we can better describe this structure as a ``reinforced bridge'', which shares the same topological placement as the other bridges, but with more elementary constituents involved rather than a single edge.


\section{Discussion}

In this section we discuss usage insights, strengths and limitations of Syndesmoscope, and directions for future work.

\subsection{De-aggregation as a Strategy}

An important design choice in Syndesmoscope invariant plots is de-aggregation. Previous invariant plots related to eccentricity and sparsity, BMatrix~\cite{Bagrow2019} and Graph Thumbnails~\cite{Yoghourdjian2016}, respectively, fully aggregated node-level information and thus make it  unavailable in their visual encodings. In contrast, kSnakes and HopCensus preserve individual node marks, while also capturing topological information in their geometric layout. With individual node marks visible and selectable, the leapfrogging and hopscotching interactions can operate at the granularity of elementary constituents, rather than of aggregates.

\subsection{Usage Insights}

Through usage scenarios we demonstrate how Syndesmoscope reveals network patterns inaccessible through any single view alone. The key insight is that a network pattern obscured or conflated in one view can surface as a legible visual pattern in another. The leapfrogging and hopscotching interaction idioms operationalize this principle: a selection initiated from a visual pattern in one pane propagates to all others, allowing users to observe whether the corresponding node and edge sets remain salient under a different, and crucially interpretable, topological axis.

In the 'Fullerene (C720)' example (Fig.~\ref{ScenarioSubgraphs}, left side), this structural insight is achieved entirely through eccentricity-based selection and hopscotching traversal. If only a ForceDirected view without this highlighting were available, these motifs would not be salient, because there would be no visual cues about this shared and recurring structural equivalence. A user would have difficulty identifying and selecting all of these subgraphs as a group. In contrast, the HopCensus pane visual pattern collapses structurally equivalent nodes into overlapping polylines within the same vertical bin, with the same eccentricity and degree dispersion. This isolation of an invariant topological property makes the selection of recurring topological patterns trivial. In Fig.~\ref{ScenarioSubgraphs} (left side), a single brush selection of the longest polylines in the bottom HopCensus bins isolates the 240 nodes with the highest geodesic eccentricity. Hopscotching their shared edges and leapfrogging to the ForceDirected pane reveals the recurring motifs in that view, through the highlighted selected nodes and edges that are distributed across many non-adjacent spatial regions of the molecule.

Consider what would be visible in the 'Western Power Grid' example (Fig.~\ref{ScenarioDensity}, without the rest of Syndesmoscope, with only the Node-Link view in the ForceDirected pane as with previous general network visualization tools (Sec.~\ref{Sec:GeneralNetworkVis}). Without Syndesmoscope, the densest $k$-5 shell is visually buried and we would have no reason to suspect anything unusual; that is, there would be no cue that this topology has a decoupled eccentricity and sparsity. However, with Syndesmoscope, selecting  the low-eccentricity polylines in HopCensus and leapfrogging to kSnakes immediately reveals they map to sparse shells: the opposite of what the axes' shared valence would predict. The topological anomaly is visually obvious because two distinct topological lenses are side by side, demonstrating the power of linking traditional views to invariant plots.

\subsection{Limitations and Future Work}

We present a preliminary qualitative analysis of the resulting visual patterns as an initial step toward understanding the potential utility of Syndesmoscope. Looking ahead, we plan to further evaluate Syndesmoscope through case studies with domain experts and, where appropriate, formal user studies. These evaluations may help identify more grounded, user-driven directions for future research. In a similar vein, empirical assessment of the kSnakes invariant plot through targeted user studies remains an open avenue for future work.

As a proof-of-concept, Syndesmoscope currently faces scalability limitations. The system architecture has not yet been optimized to process and render very large datasets, and the static visual encodings of kSnakes and HopCensus can become visually cluttered as network size increases, as observed in some examples in the Appendix. Future work may explore visual adaptations of these invariant plots to better accommodate larger topologies.

From a technical perspective, one possible direction is the development of a runtime backend that allows users to directly load CSV topology datasets, with the system computing the corresponding JSON geometric layouts and rendering the results on demand. Such an architecture could also support interactive topological editing, enabling users to add or remove nodes and edges, further extending the set of interaction idioms. Additionally, a metadata passthrough mechanism could preserve data attributes during isomorphic randomization, allowing semantic information to appear alongside invariant visual patterns.


\section{Conclusion}

The structural diversity of graph topology is difficult to capture through any single algorithmic lens or network visualization idiom. Syndesmoscope addresses this challenge by juxtaposing invariant plots and traditional views, with linked highlighting across visual patterns along interpretable axes, which enable the exploration of topology through different interpretable algorithmic lenses at once. With the interactive affordances of leapfrogging across visual patterns and hopscotching, users can intuitively explore patterns from mesoscale phenomena.


\section{Supplemental Materials Index}

The Appendix includes additional usage scenarios. The Supplemental materials include further details on isomorphic randomization and Census-Stub pre-seriation (\SuppOne), shared index JSON data structure schemas (\SuppTwo), examples of Syndemoscope's responsive portrait mode (\SuppThree), and details about the corpus and dataset provenance (\SuppFour). Supplemental materials, including an archived codebase, are available at: https://osf.io/puk8a. \textbf{Live demo:} https://syndesmoscope.vercel.app. \textbf{Source code:} https://github.com/dirediredock/Syndesmoscope/releases/tag/v1



\acknowledgments{The development of this paper took place between October 2025 and March 2026 at the UBC Point Grey campus and in the City of Vancouver, which sit on the traditional, ancestral, unceded territory of the Musqueam, Squamish, and Tsleil-Waututh First Nations. This work was supported in part by NSERC DG RGPIN-2024-06401. We thank Christopher Mundle and Peyton Rapo for software engineering advice; and Ricky Curry, Steve Kasica, Ryan Smith, and Mara Solen for manuscript reviews. We used the AI-assistance tools Claude Code and Replit Agent 4 for frontend deployment.}




\appendix

\section{Appendix}
\label{Sec:Appendix}

In this Appendix we present 12 more scenarios, in addition to the 6 usage scenarios in the main paper. 

\begin{figure*}[!t]
\centering
\includegraphics[width=0.92\linewidth]{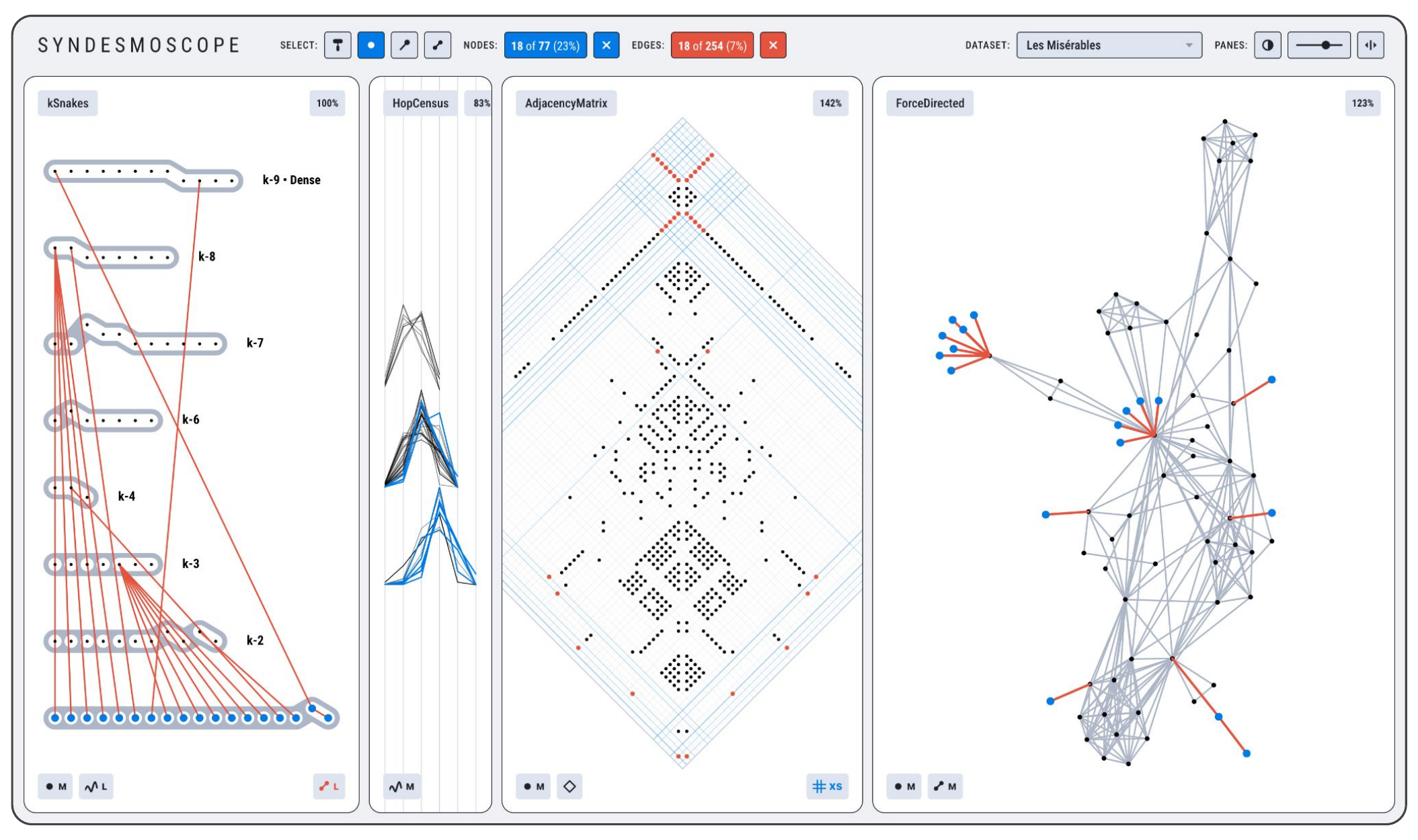}
\caption{The 'Les Misérables' literary dataset: 77 characters (nodes) and 254 spoken interactions (edges). In the kSnakes pane, we brush-select all the nodes in the sparsest shell (bottom of the pane), then we hopscotch their stubs to reveal where these links appear in the ForceDirected node-link view. In the HopCensus pane these selected nodes appear at the bottom of the pane so we see they have high eccentricity, while in the AdjacencyMatrix pane the placement of the edges near both poles, rather than  near the equator. In the kSnakes pane, the edge overlays include two edges that span the entire dense-sparse gradient of the graph, connecting two of the leaf nodes in the sparsest kSnakes shell to nodes in the densest shell at the top of the pane.}
\label{APX_kSnakes_01}
\end{figure*}

\begin{figure*}[!t]
\centering
\includegraphics[width=0.92\linewidth]{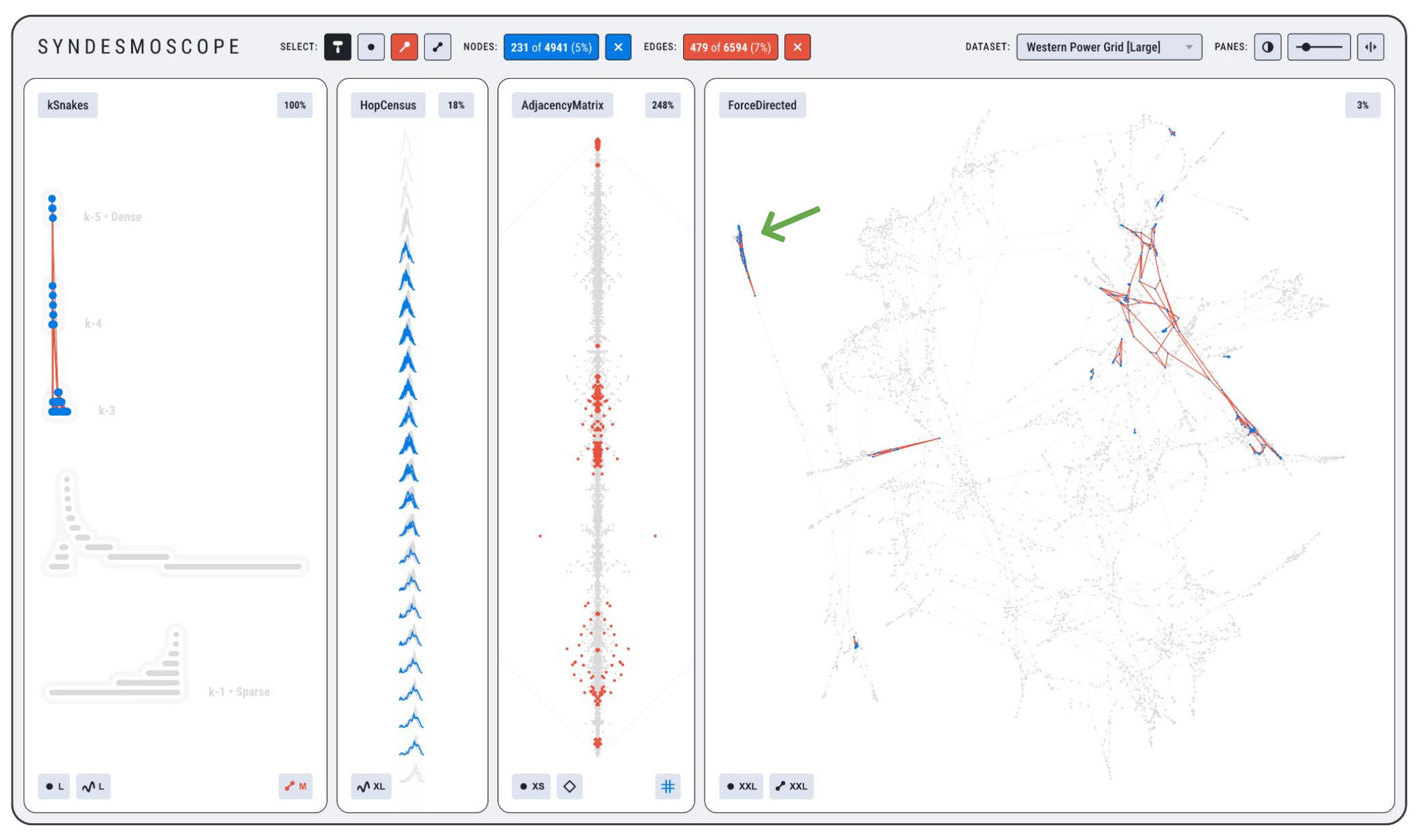}
\caption{The 'Western Power Grid' infrastructure dataset: 4941 substations (nodes) and 6594 transmission lines (edges). We continue the investigation begun in Fig.~\ref{ScenarioDensity}, by selecting the three most dense shells at the top of the kSnakes pane. We see that these selected shells span many levels of eccentricity in the HopCensus pane. High eccentricity values are counterintuitive for nodes in dense shells. Further use of Syndesmoscope revealed that the densest $k$-5 shell of the graph, as shown by the green arrow annotation in the ForceDirected pane, appears in the topological perimeter, which is the phenomenon that leads to skewed overall eccentricity lengths.}
\label{APX_kSnakes_02}
\end{figure*}

\begin{figure*}[!t]
\centering
\includegraphics[width=0.92\linewidth]{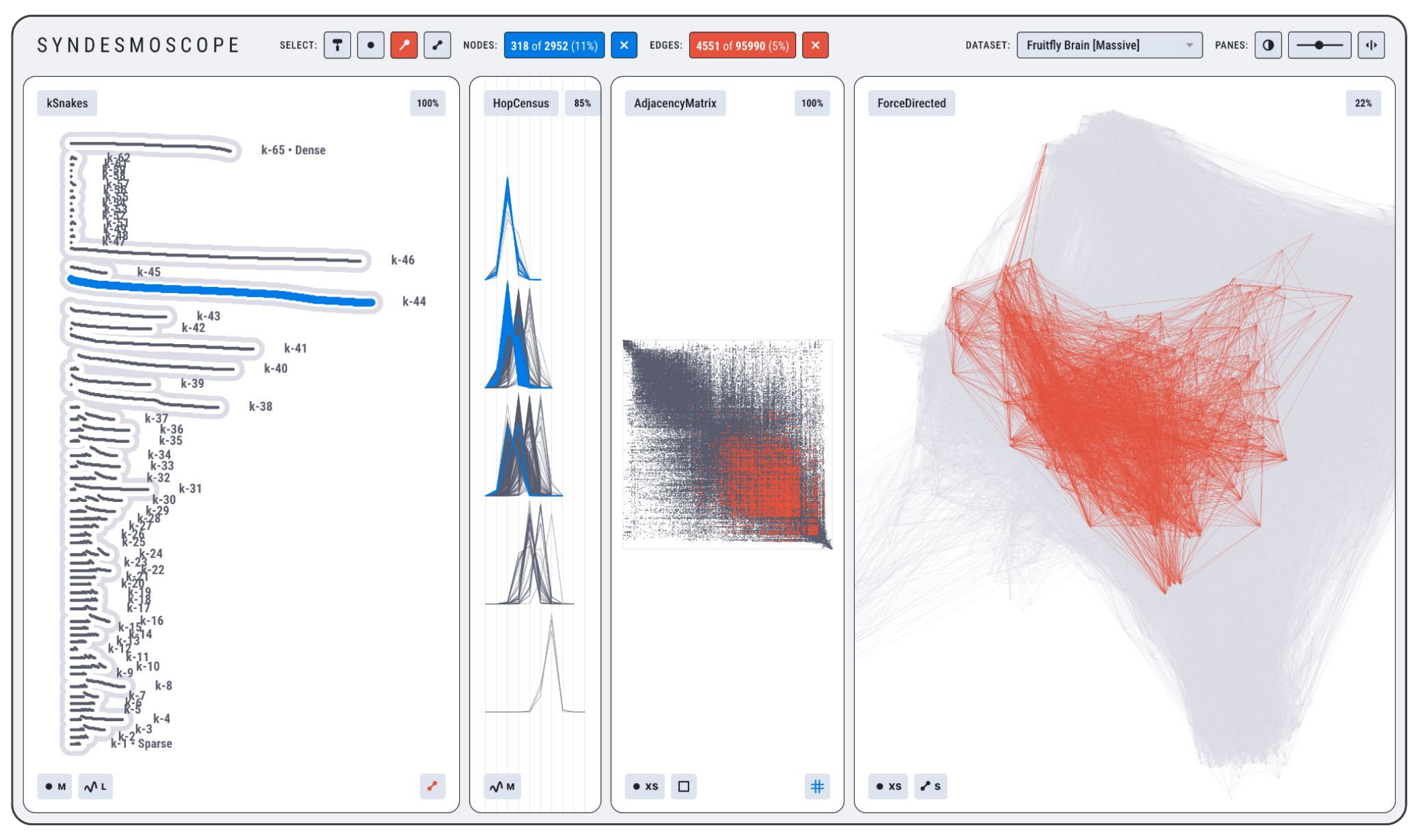}
\caption{The 'Fruitfly Brain' neurological dataset: 2956 neurons (nodes) and 116922 synapses (edges). This large graph shows the scalability limits of Syndesmoscope. The kSnakes pane is very crowded, and the annotations of the per-shell $k$ values overlap vertically, and the large number of nodes in some shells causes them to overlap horizontally. Nonetheless, the shells across the vertical dense-sparse gradient are salient and selection of node points within shells is still possible. We can see that the selected $k$-44 shell nodes range across a large section of the ForceDirected pane, from the middle out towards the border.}
\label{APX_kSnakes_03}
\end{figure*}


\begin{figure*}[!t]
\centering
\includegraphics[width=0.92\linewidth]{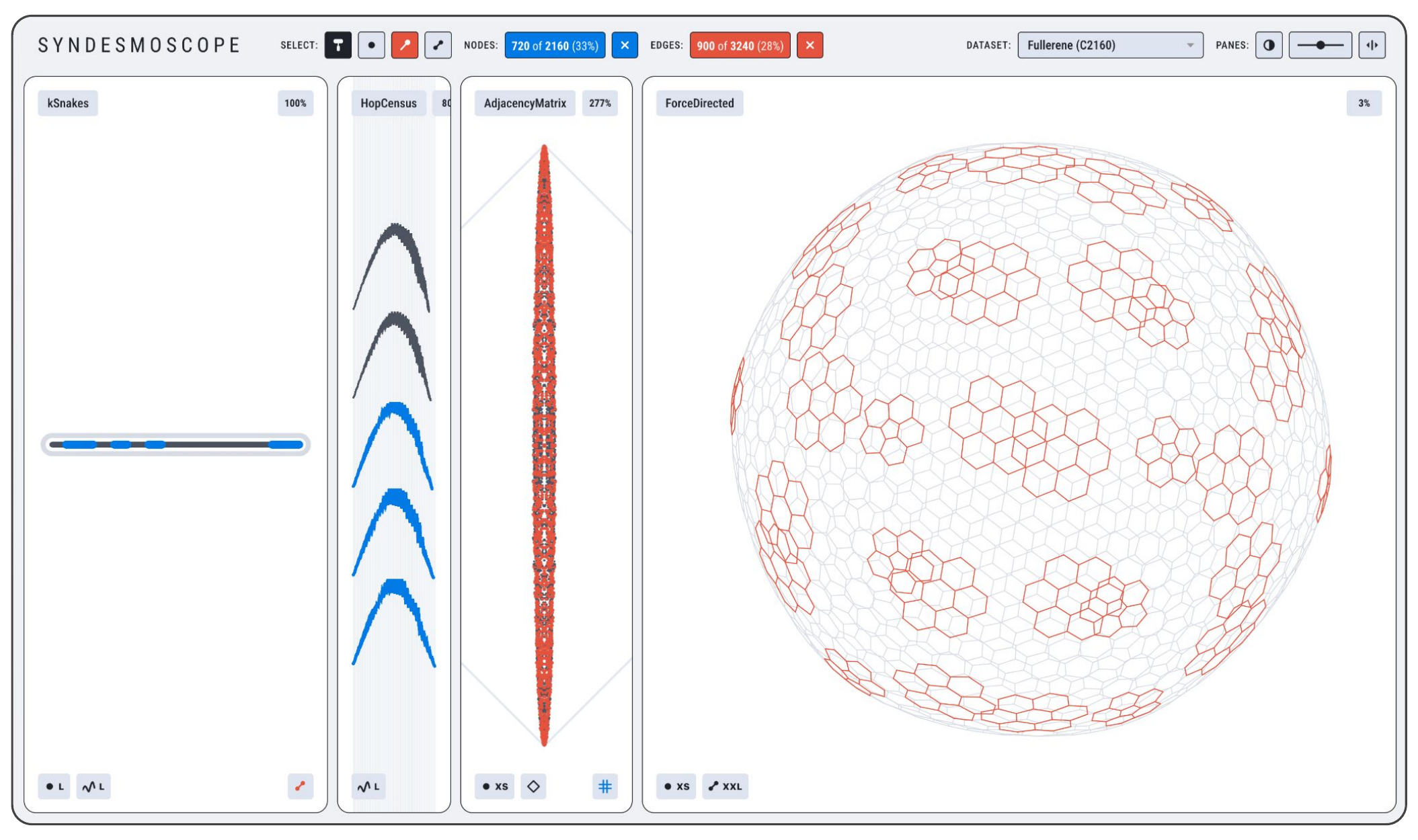}
\caption{The 'Fullerene (C2160)' geometric dataset: 2160 carbon atoms (nodes) and 3240 bonds (edges). Selection of the high eccentricity polylines at the bottom of HopCensus reveals recurring pentagonal and hexagonal "floret" motifs in the molecule. Here we show the same selection process as in Fig.~\ref{ScenarioSubgraphs} (left side) of the main paper, but with a larger-scale member of the Fullerene family. In this case we can see that there is no variance in the visual pattern of kSnakes, as all nodes belong to the same shell and have the same onion value. Similarly, the AdjacencyMatrix view does not show a useful pattern because the recurring motifs do not align with the graph's bisection axis.}
\label{APX_HopCensus_01}
\end{figure*}

\begin{figure*}[!t]
\centering
\includegraphics[width=0.92\linewidth]{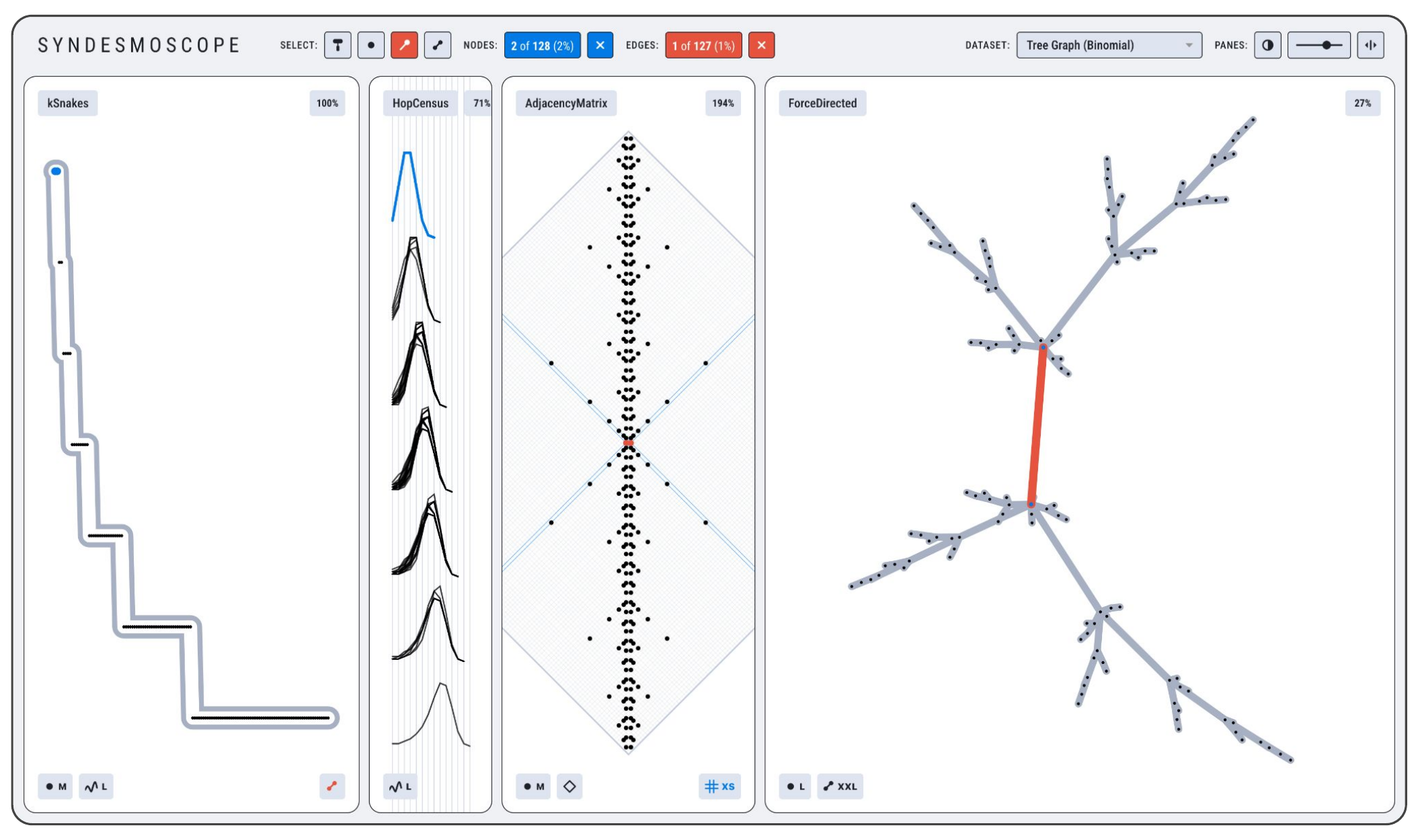}
\caption{The 'Tree Graph (Binomial)' simple dataset: 128 nodes and 127 edges. In the HopCensus pane, the shortest polylines at the top, which represent the central nodes, are selected. As removing any edge in a tree disconnects the graph, all edges in a tree are bridge edges. With Syndesmoscope we can find the "edgiest edge", which in this case is the edge that, if removed, separates the tree into two halves. We can see the interesting nature of that edge in the ForceDirected view in this case, but we can confirm its nature by leapfrogging to the other panes to understand more about its properties. Its nodes have the highest onion values, as we see at the top of the kSnakes pane, and are found at the exact equator of the bisection axis, as we see in the AdjacencyMatrix pane.}
\label{appendix/APX_HopCensus_02}
\end{figure*}

\begin{figure*}[!t]
\centering
\includegraphics[width=0.92\linewidth]{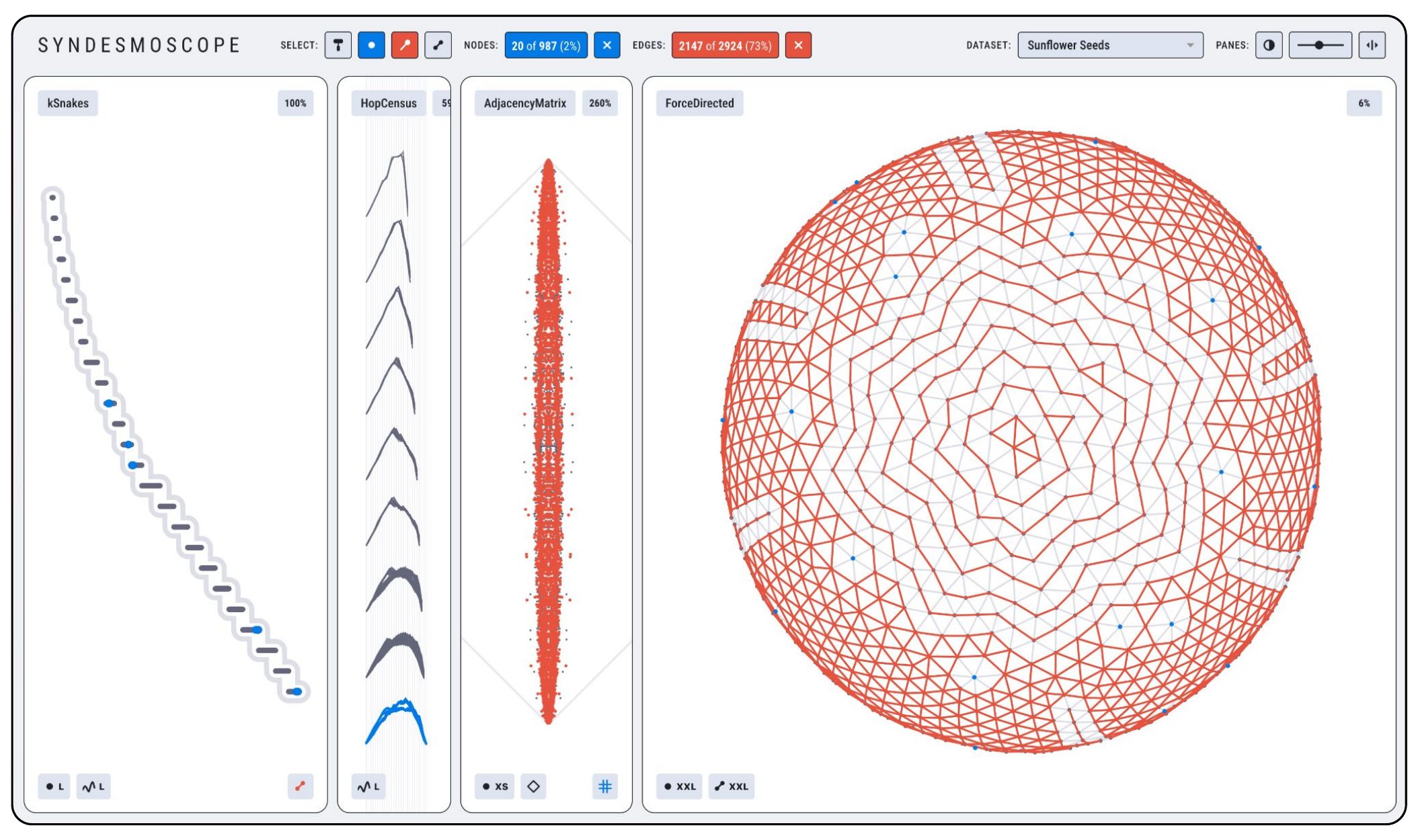}
\caption{The 'Sunflower Seeds' geometric dataset: 987 seeds (nodes) and 2924 neighbor connections (edges). This image is the result of a complex series of selection actions. We first select a bin in the HopCensus plot, then hopscotch to its shared edges, and then clear the node selection set while leaving the edges selected. We then repeat this process for all other bins in HopCensus. The resulting set of selected edges consists of contours of equal-eccentricity edges in the ForceDirected pane, revealing interesting mesoscale structure in the arrangement of sunflower seeds.}
\label{appendix/APX_HopCensus_03}
\end{figure*}


\begin{figure*}[!t]
\centering
\includegraphics[width=0.92\linewidth]{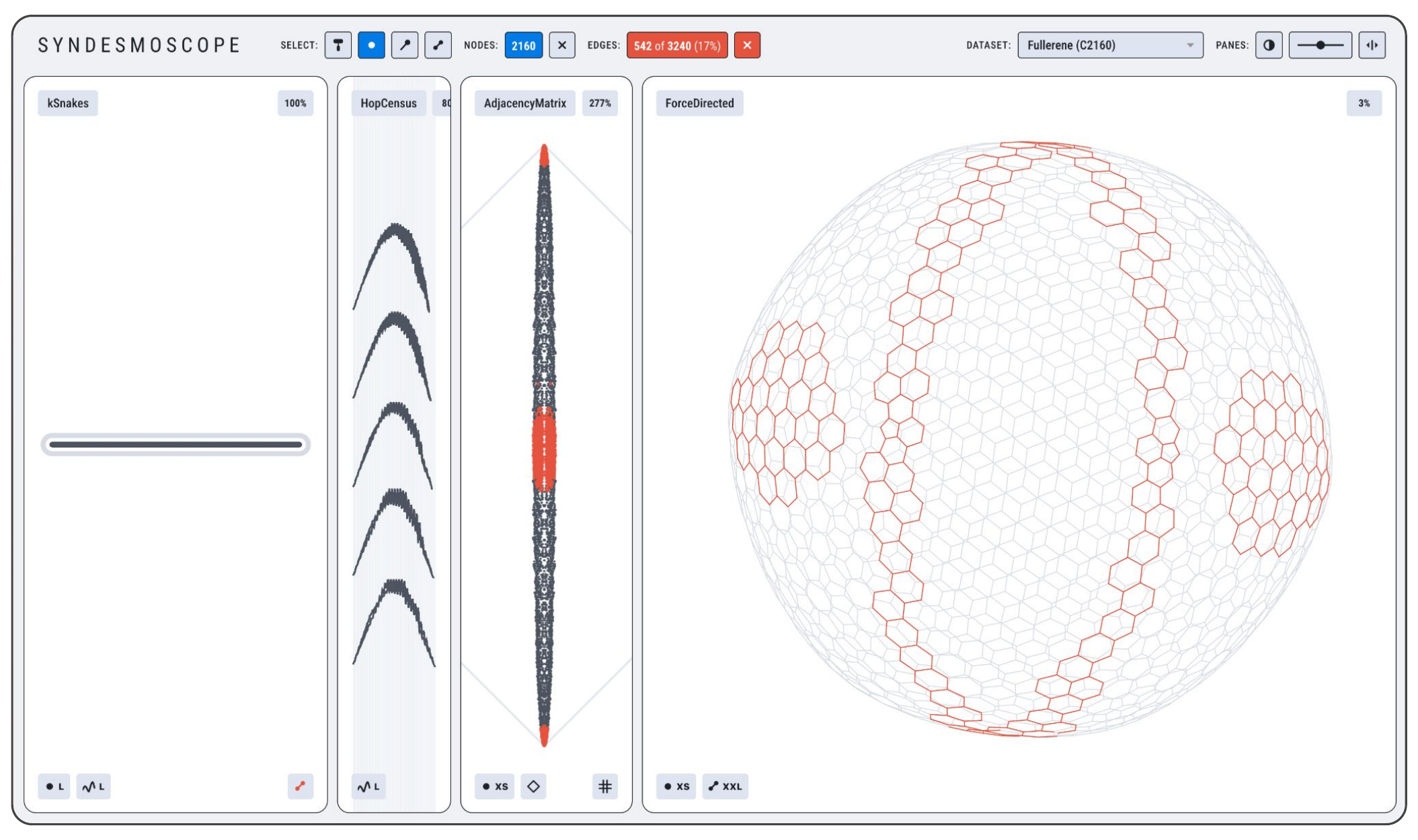}
\caption{The 'Fullerene (C2160)' geometric dataset: 2160 carbon atoms (nodes) and 3240 bonds (edges). With the diamond orientation of the AdjacencyMatrix pane, selection of several equal-sized areas that occur in different regions along the bisection axis from poles to equator is easy. In the ForceDirected pane, we see bands around the spherical shape that are parallel to each other.}
\label{APX_AdjacencyMatrix_01}
\end{figure*}

\begin{figure*}[!t]
\centering
\includegraphics[width=0.92\linewidth]{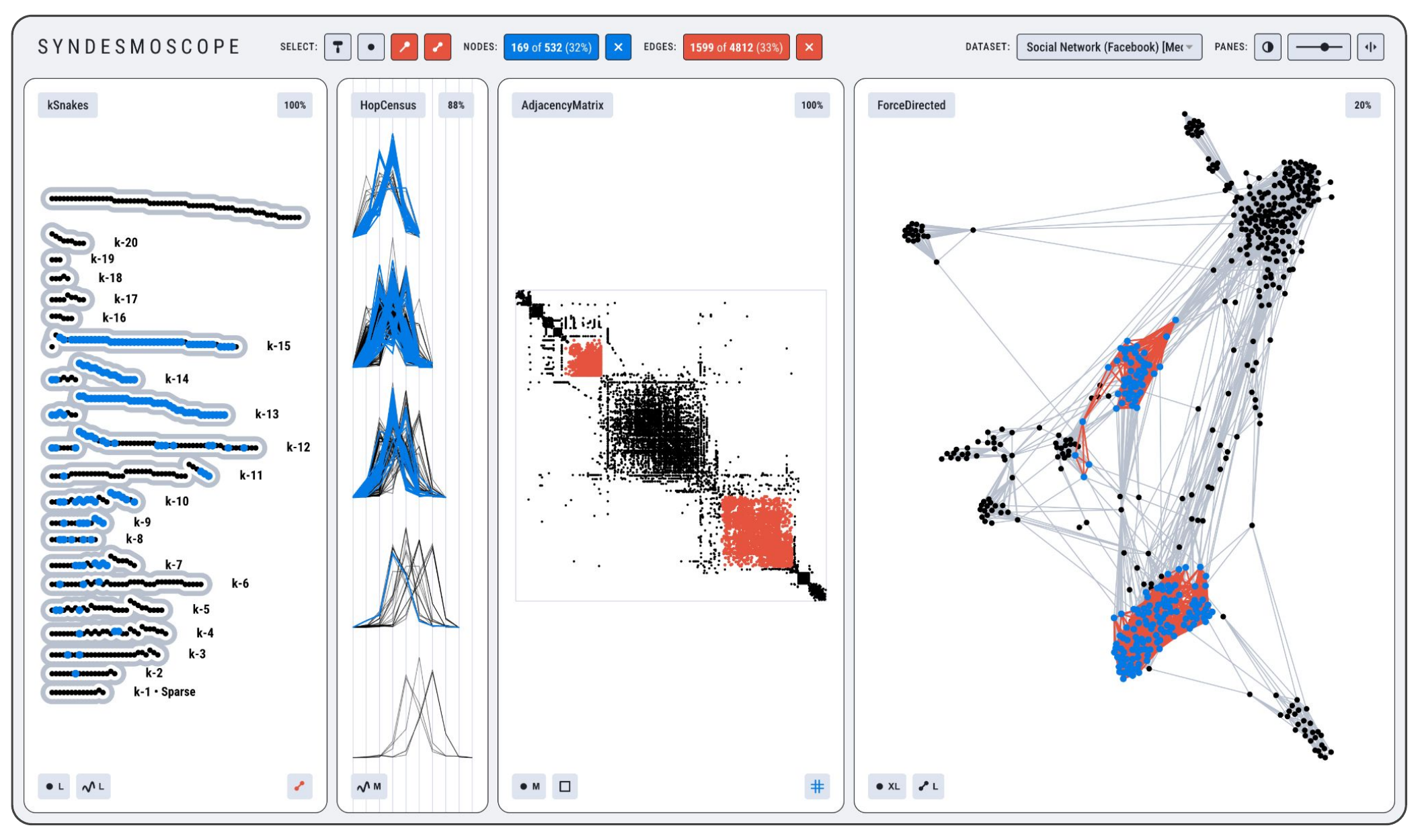}
\caption{The 'Social Network (Facebook)' social network dataset: 2160 people (nodes) and 3240 friendships (edges). With the square orientation of the AdjacencyMatrix pane, selection of the rectangular clusters in the visual pattern is easy. In this example, two clusters that are equidistant from the bisection equator are selected, and we can see how they map to visual clusters in the ForceDirected pane. While these clusters have central nodes, as seen in short polylines at the top of the HopCensus pane, we can see that none of these clusters have dense nodes because nothing is highlighted at the top of the kSnakes pane.}
\label{APX_AdjacencyMatrix_02}
\end{figure*}

\begin{figure*}[!t]
\centering
\includegraphics[width=0.92\linewidth]{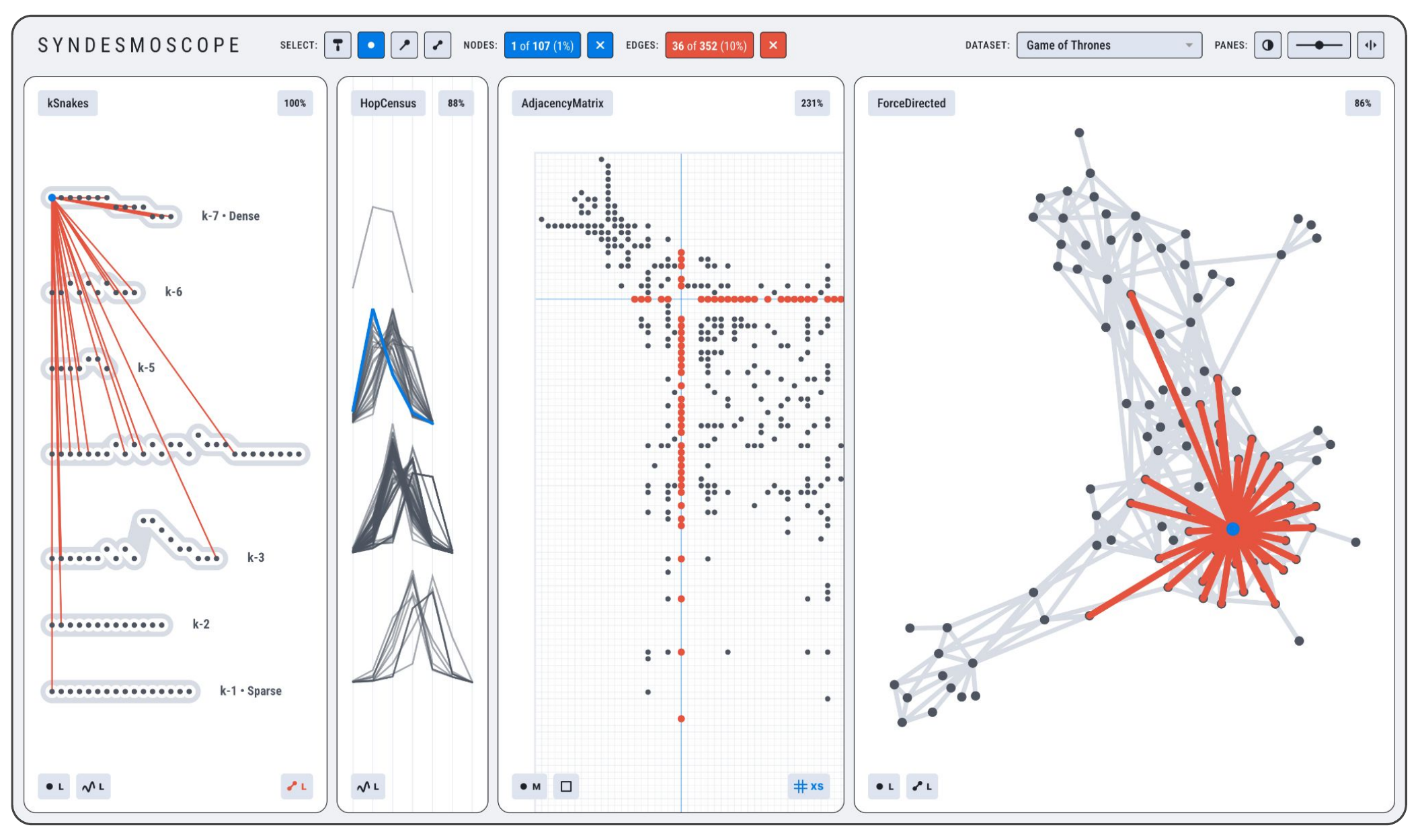}
\caption{The 'Game of Thrones' literary dataset: 107 book characters (nodes) and 352 spoken interactions (edges). We can find the highest degree node by exploring with the combination of the Syndesmoscope panes. Selection of potential candidates is easiest in the Adjacency Matrix pane by using the square orientation in conjunction with a very thin brush, and then noticing the edge count in the control strip on top. We can also check if the node appears in the most dense shell in the kSnakes view, and cross-check it in the ForceDirected pane. In this example, we have selected the highest degree node, and can see that it is nearly central from the HopCensus pane, because it is in the second from the top bin. Hopscotching to its stubs reveals the reach of this high-degree node, which spans all shells in the dense-sparse gradient of the kSnakes pane.}
\label{APX_AdjacencyMatrix_03}
\end{figure*}


\begin{figure*}[!t]
\centering
\includegraphics[width=0.92\linewidth]{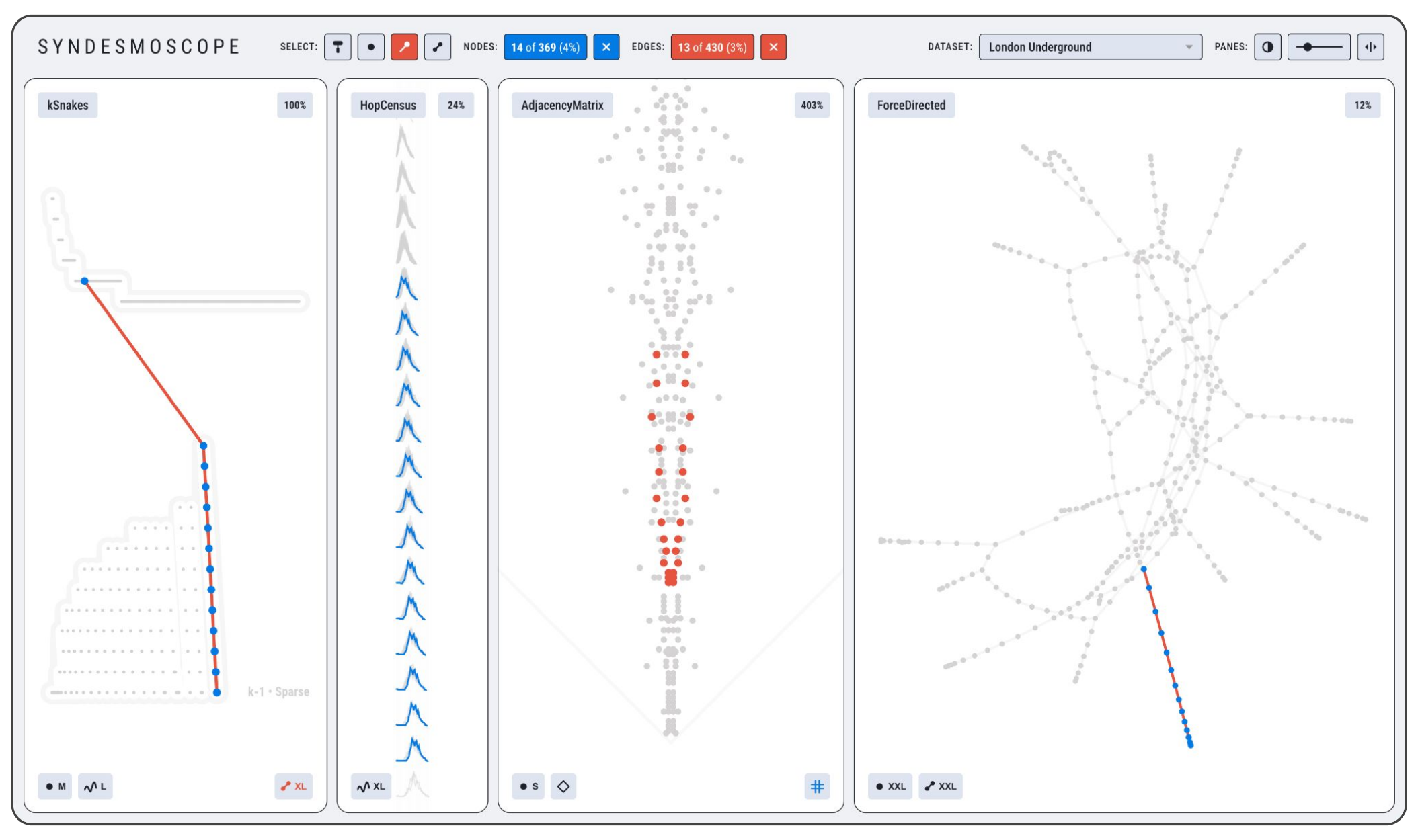}
\caption{The 'London Underground' infrastructure dataset: 369 stations (nodes) and 430 connections (edges). We select a long chain in the ForceDirected pane, and discover by leapfrogging to the kSnakes view that it is the longest chain in the sparsest shell. As expected, many eccentricities are highlighted in HopCensus, because a chain has monotonically distant nodes in terms of hops, each with one eccentricity value higher than the one before it in the chain.}
\label{APX_ForceDirected_01}
\end{figure*}

\begin{figure*}[!t]
\centering
\includegraphics[width=0.92\linewidth]{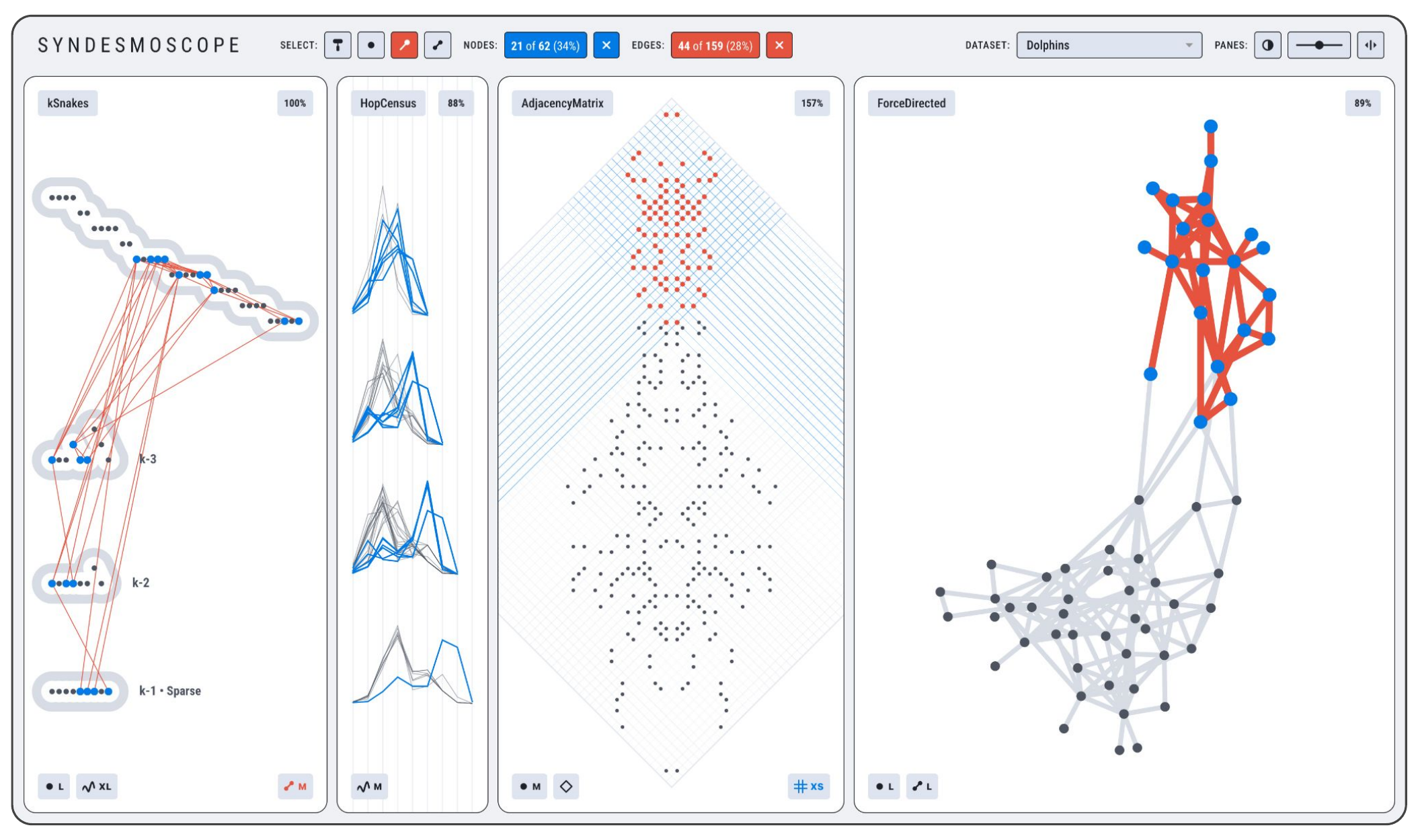}
\caption{The 'Dolphins' ecological dataset: 62 dolphins (nodes) and 159 interactions (edges). A brush selection of the upper half in the ForceDirected pane reveals that these nodes and edges belong to only one pole of the AdjacencyMatrix bisection axis. In this small graph, the nodes and edges span all levels of the eccentricity and dense-sparse gradients.}
\label{APX_ForceDirected_02}
\end{figure*}

\begin{figure*}[!t]
\centering
\includegraphics[width=0.92\linewidth]{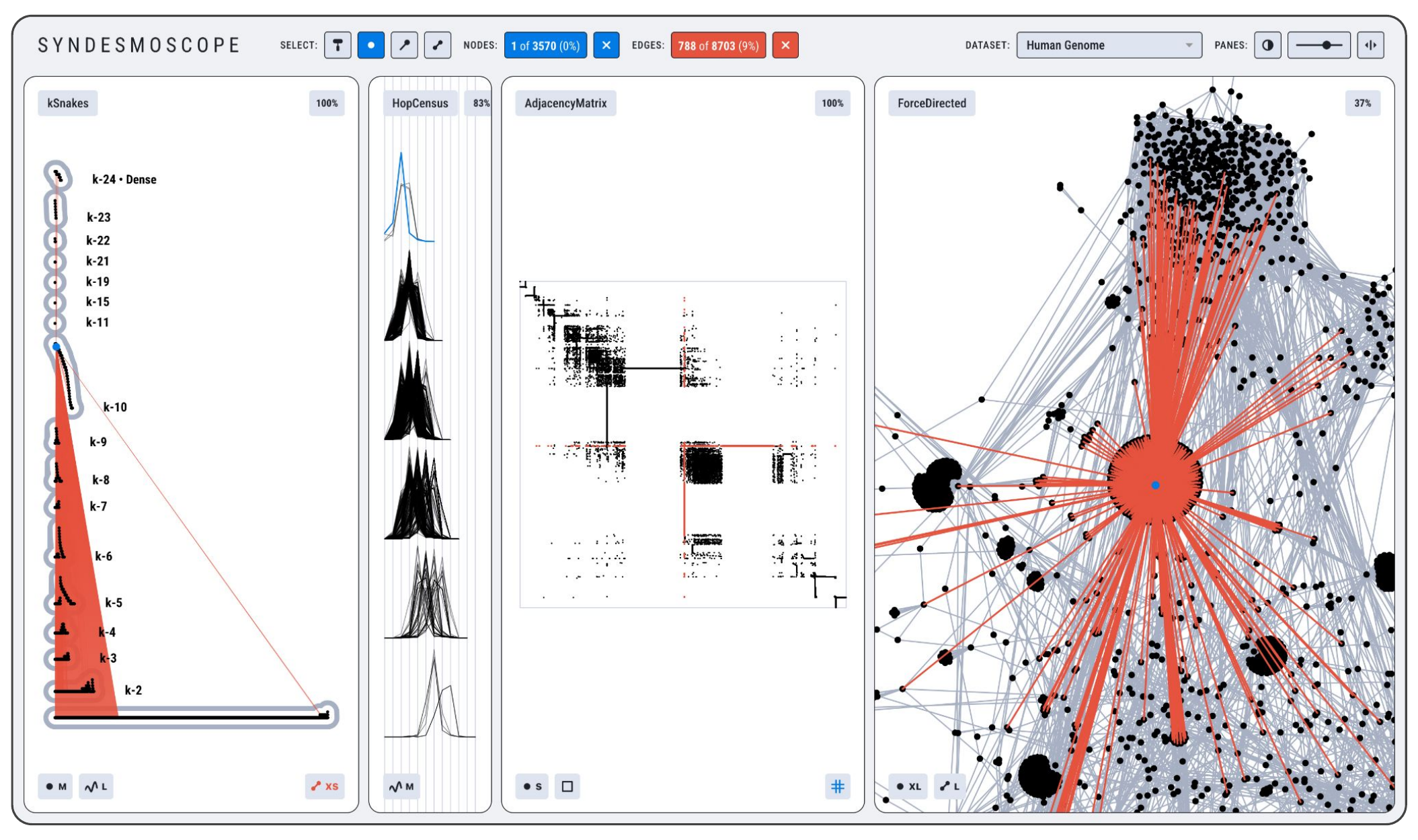}
\caption{The 'Human Genome' biological dataset: 3751 genes (nodes) and 9034 genetic interactions (edges). We select a high-degree node in the ForceDirected pane, and hopscotch via its stubs to inspect its one-hop neighborhood in that view. Leapfrogging to the HopCensus pane reveals that this node is central because it is in the topmost bin, while the kSnakes pane shows that this node has the highest onion value of all the nodes in its $k$-10 shell. We can also see most of the direct connections to this node are leaves, because they appear in the bottom-most shell.}
\label{APX_ForceDirected_03}
\end{figure*}


\end{document}